\documentclass[12pt]{article}
\usepackage{epsfig}
\newcommand{\beq}{\begin{equation}}
\newcommand{\eeq}{\end{equation}}
\newcommand{\0}{x \,}
\newcommand{\1}{x_1 \,}
\newcommand{\2}{x_2 \,}
\newcommand{\3}{x_3 \,}
\newcommand{\4}{x_4 \,}

\newcommand{\bl}{\bigl\langle\,}
\newcommand{\br}{\,\bigr\rangle}

\newcommand{\dd}{\partial}

\newcommand{\di}{\displaystyle}
\newcommand{\la}{\lambda}

\newcommand{\Om}{\Omega}

\newcommand{\ve}{\varepsilon}
\newcommand{\vp}{\varphi}

\newcommand{\A}{\frac{1}{8\pi^2}\ln\left(\frac{m^2 e}{\Lambda^2}\right)-J_0(m,T)}
\newcommand{\vecp}{\textit{\textbf{p}}}
\newcommand{\veck}{\textit{\textbf{k}}}
\newcommand{\vecx}{\textit{\textbf{x}}}

\newcommand{\s}{_{sym}}

\title{
Bogolubov's chain of equations method for temperature Wightman
functions in thermodynamics of relativistic  phase transition.}

\author{G.M. Vereshkov O.D. Lalakulich A.V. Kartavtsev\footnote{e-mail: akartavt@uic.rnd.runnet.ru}}

\date{Research Institute of Physics, Rostov State University, Russia}

\begin{document}
\maketitle
\begin{center}
\fontsize{10}{10}\selectfont
\parbox[c]{140mm}{Bogolubov's chain of equations method for temperature
Wightman functions is suggested for investigation of relativistic
phase transition. The chain equations for the Wightman functions
forming momentum--energy tensor are obtained. It is clarified that
structure  of the chain equations determines the basis
approximation (the Hartree - Fock approximation)  and corrections
calculation algorithm.  The basis approximation is investigated in
details: renormalized equations for effective masses, order
parameter and generating functional which reproduce those
equations are obtained. Being considered on the solution of the
gap equations  for the effective masses the generating functional
turns to nonequilibrium  functional of free energy density, which
allows to obtain phases stability conditions. Thermodynamic
observables like heat capacity and sonic speed are calculated. The
correction to the Hartree-Fock approximations is ascertained to be
small for all temperatures excluding vicinity of the phases
equilibrium point.}\end{center}

\section{Introduction.}
An idea that phase transition may take place in a quantum field
system was first proposed by Kirznits\cite{kirznits1}. First
quantitative calculations were made by Kirznits and Linde
\cite{KirLinde1}, and later by Dolan and Jackiw \cite{DolJackiw},
Weinberg \cite{weinberg}. Theory of relativistic phase transition
(RPT) deals with nonequilibrium Landau functional (which is
usually called ``effective potential''). When quasi--particle
approximation is used for calculation of effective potential  from
microscopic operational Lagrangian, a problem of quasi--particles
mass spectra arises. Calculations made in Matsubara diagram
technique showed, that phase transitions can't be described in
iterative perturbation theory approach --- tachion pole in mass
spectra of scalar particles and consequently imaginary terms in
Matsubara ``effective potential'' appears far from the point of
phase transition. Enhanced resuming schemes successfully solved
the problem of tachion pole in mass spectra but not the problem of
effective potential calculation.

As for calculation of nonequilibrium functional in Matsubara
approach, with temperature Green functions obtained in the
framework of resuming schemes, the extremum condition of such
functional leads to equation for order parameter which is
different from that obtained directly from equations of motion.
Illusion of solving of the problem mentioned is achieved only by
manipulation with high-temperature approximation of Bose-Einstein
integrals.

Cornwall, Jac\-kiw and Tomboulis suggested an ``effective
functional'' of different kind \cite{CJT}. In addition to equation
for order parameter the "CJT-functional" reproduces equations for
full Green functions. At present $O(N)-$model is almost completely
investigated in this approach in Hartree-Fock approximation.
Nevertheless problems of renormalization of the functional,
analytical analysis of the phases stability conditions,
calculation of observables and estimation of Hartree-Fock
approximation accuracy need to be discussed in details.

These problems (at least for $O(N)-$model) can be solved in
alternative approach, based on Bogolubov's chain of equations for
temperature Wightman functions (WF).

In Wightman approach the equations of motion and observables of a
quantum--field system are written in terms of Wightman functions.
\cite{wightman}. In this paper equations for all the functions
forming energy-momentum tensor are presented. Field operators
smoothing, used in axiomatic Wightman formalism, replaced with
scheme of dimensional regularization. Self--consistency of that
redefinition method is proved.

It is found out that effective masses of quasi--particles, which
are formed by particles interactions with each other and vacuum,
automatically arise in the Bogolubov's chain equations. Thus, it
is natural to use Hartree--Fock approximation as a basis
approximation of reduction. Corrections calculation algorithm is
also determined by the Bogolubov's chain mathematical structure.

Hartree--Fock approximation was investigated in details. A
functional (generating functional) which reproduce gap equations
for effective masses and equation for order parameter is obtained.
It is shown, that the generating functional can also be calculated
from momentum--energy tensor components, which proves
self--consistency of thermodynamic equations of state obtained
from quantum equations of motion with those obtained from
generating functional. The generating functional, being considered
on solutions of the gap equations for effective masses turns into
functional of nonequilibrium free energy, which is used for
analytical analysis of the phases stability condition. Functional
of equilibrium free energy, which is obtained by substitution of
the order parameter on temperature dependence into nonequilibrium
one, is used for calculation of thermodynamical observables: heat
capacity and sonic speed.

An important problem of the  Hartree--Fock approximation validity
was clarified by calculation of correction of the first order of
vanishing to this approximation.

A Lagrangian of the $O(N)-$model and equations of motions are
written out in chapter \ref{O(N)Lagrangian}. A definition of
Wightman functions, symmetrized with regard to permutation of it's
arguments, chain's equations for Wightman functions up to fourth
rank and chain reduction algorithm are given in chapter
\ref{Tcepochka}. Equations for effective masses and order
parameter, functionals and observables in Hartree--Fock
approximation, and results of numerical calculations are given in
chapter \ref{PSP}. Finally, chapter \ref{Popravka} is devoted to
calculation of the first order of vanishing corrections to
effective masses, order parameter and observables.

\section{Lagrangian and equations of motion.\label{O(N)Lagrangian}}
Operator equations of motions for quantum fields ore obtained from
Lagrangian of system of $N$ scalar fields $\phi_a$
\begin{equation}
{\cal L}=\frac12\left(\partial_{\mu} \phi_a\cdot\partial^{\mu}
\phi_a+ \mu^2\phi_a\cdot\phi_a\right)
-\frac{\la}{N}(\phi_a\cdot\phi_a)^2, \label{lagrangian}
\end{equation}
\noindent by standard variational procedure:
\[
\dd_{\mu}\dd^{\mu}\phi_a-\mu^2\phi_a+\frac{4\la}{N}(\phi_b\phi_b)\phi_a=0.
\]
For $\mu^2<0$, the theory is invariant under  $O(N)$
transformations of the quantum fields. For $\mu^2>0$ and low
temperatures spontaneous symmetry breaking described by nonzero
vacuum expectation value of $\phi_N$,
\[
\langle |\phi_N |\rangle=v
\]
breaks down symmetry group to $O(N-1)$ and leads to $N-1$
Goldstone bosons.
\[
\begin{array}{l}
\phi_N\equiv\vp+v, \\
\phi_a\equiv\chi_a,\hspace*{2mm}a=1\ldots N-1
\end{array}
\]
Equations for quantum fields $\chi_a$ and $\vp$ look like:
\begin{equation}
\dd_{\mu}\dd^{\mu}\vp-\mu^2(v+\vp)+\frac{4\la}{N}(v^3+3v^2\vp+3v\vp^2+\vp^3
+v\chi_a \chi_a+\vp\chi_a \chi_a)=0 \label{vp}
\end{equation}
\begin{equation}
\dd_{\mu}\dd^{\mu}\chi_a-\mu^2\chi_a+\frac{4\la}{N}(v^2\chi_a+2v\vp\chi_a+\vp^2\chi_a+\chi_b
\chi_b\chi_a )=0 \label{chi}
\end{equation}

Momentum--energy tensor is obtained from Lagrangian
(\ref{lagrangian}) by metric variation procedure. For homogeneous,
isotropic and stationary system, momentum--energy tensor turns to:
\begin{equation}
\begin{array}{c}
\di \langle T_ \mu^\nu\rangle=\langle
\partial_\mu\vp\partial^\nu\vp\rangle+
\langle\partial_\mu\chi_a\partial^\nu\chi_a\rangle-
\\[2mm]\di
-\delta_\mu^\nu\biggl[\frac{\mu^2 v^2}{2}-\frac{\la
v^4}{N}+\frac{\la}{N}\biggl(\langle\vp^4\rangle+\langle\chi_a
\chi_a\chi_b\chi_b
\rangle+2\langle\vp^2\chi_a\chi_a\rangle+2v\langle\vp^3\rangle+2v
\langle\vp\chi_a\chi_a\rangle \biggr)\biggr]. \label{TofEI}
\end{array}
\end{equation}

\section{Bogolubov's chain of equations.\label{Tcepochka}}
\paragraph{Definition of symmetrical Wightman functions.}
Dynamics of a system, which is determined by Lagrange equations of
motions, can be described in terms of chain of equations for
Wightman functions (WF) --- so called Bogolubov's chain.

Full $n-$point WF (WF of rank $n$) is an over state expectation
value of $n$ quantum field operators, taken, generally speaking,
at different space--time points.
\[
\bl\vp_{(\1)}\vp_{(\2)}..\vp_{(x_l)}...\chi_{(x_m)}..\chi_{(x_n)}
\br.\] It can be represented as a sum of various WF of lower rank
(the sum of ranks in each product is equal to $n$) and so called
correlative Wightman function of the same rank. This
representation, essentially, is a definition of correlative WF.
For homogeneous, isotropic space--time it is convenient to use WF
symmetrized with regard to permutation of it's arguments.
Algorithm of it's definition, which takes into account residual
$O(N-1)$ symmetry of the theory under consideration, is
illustrated for two--point $G_{(\1\2)}$, $D_{(\1\2)}$ and
four--point $G_{_{22} (\1\2|\3\4)}$ functions. Here left bottom
index is the number of $\vp$ fields and the right one is the
number of $\chi_a$ fields which form the corresponding Wightman
function.

1. An expectation value of a sum of every possible permutation of
quantum--field operators is taken (symmetrization operation):
\[
\begin{array}{c}
\hspace*{0mm}2G_{(\1\2)}\equiv\bl\vp_{(\1)}\vp_{(\2)}\br\s
\equiv\bl\vp_{(\1)}\vp_{(\2)}\br+\bl\vp_{(\2)}\vp_{(\1)}\br ,\\[2mm]
\hspace*{0mm}2(N-1)D_{(\1\2)}\equiv\bl\chi_{a (\1)}\chi_{a
(\2)}\br\s
\equiv\bl\chi_{a (\1)}\chi_{a (\2)}\br+\bl\chi_{a (\2)}\chi_{a (\1)}\br ,\\[2mm]
\hspace*{0mm}4 (N-1) G_{_{22}
(\1\2|\3\4)}\equiv\bl\vp_{(\1)}\vp_{(\2)}\chi_{a (\3)}\chi_{a
(\4)}\br\s\equiv\bl\vp_{(\1)}\vp_{(\2)}\chi_{a (\3)}\chi_{a
(\4)}\br+\\[2mm]\hspace*{0mm}+
\bl\vp_{(\1)}\vp_{(\2)}\chi_{a (\4)}\chi_{a (\3)}\br+
\bl\vp_{(\2)}\vp_{(\1)}\chi_{a (\3)}\chi_{a (\4)}\br+
\bl\vp_{(\2)}\vp_{(\1)}\chi_{a (\4)}\chi_{a (\3)}\br.
\end{array}
\]
Since WF should be $O(N-1)$ invariant, quantum--field operators
$\chi_a$ come in pairs with the same indexes (Einstein sum rule is
used).

In that way defined functions are invariant of permutation of
arguments of $\vp$ operators and of $\chi_a$ operators with the
same indexes.

2. Four--point  WF $G_{_{22} (\1\2|\3\4)}$ can be represented as
sum of various products of (non--symmetrized) WF of the second
rank and corresponding correlative function:

\[
\begin{array}{c}
4 (N-1) G_{_{22} (\1\2|\3\4)}=\bl\vp_{(\1)}\vp_{(\2)}\br\bl\chi_{a
(\3)}\chi_{a (\4)}\br+
\bl\vp_{(\1)}\vp_{(\2)}\br\bl\chi_{a (\4)}\chi_{a (\3)}\br+\\[2mm]
\bl\vp_{(\2)}\vp_{(\1)}\br\bl\chi_{a (\3)}\chi_{a (\4)}\br+
\bl\vp_{(\2)}\vp_{(\1)}\br\bl\chi_{a (\4)}\chi_{a (\3)}\br+4 (N-1)
C_{_{22} (\1\2|\3\4)}
\end{array}
\]

3. The sum terms can be regrouped, which picks out symmetrized WF:
\[
G_{_{22} (\1\2|\3\4)}=G_{(\1\2)}D_{(\3\4)}+C_{_{22} (\1\2|\3\4)}.
\]
It should be noted that permutation of arguments separated by
vertical line leave a Wightman function, in particular $G_{_{22}
(\1\2|\3\4)}$, invariant: $\di
 G_{_{22}
(\1\2|\3\4)}=G_{_{22} (\2\1|\3\4)}=G_{_{22} (\1\2|\4\3)}=G_{_{22}
(\2\1|\4\3)}$.

It is convenient to divide  WF which depend on odd number of
arguments by expectation value $v$.

Residual global and discrete ($\chi_a\to\--\chi_a$) symmetries of
the theory restrict number of nontrivial WF. In particular, there
exists only two nontrivial three--point
\[
C_{_{03} (\1 \2 \3)}=
              \frac1{3!\hspace*{1mm} v}
                  \bl \vp_{(x_1)} \vp_{(x_2)} \vp_{(x_3)} \br\s
\label{C03}
\]
 \[ C_{_{21} (\1 \2| \3)}=
              \frac1{2!\hspace*{1mm} v}\frac1{N-1}
               \bl \chi_{a(x_i)}\chi_{a(x_j)}\vp_{(\3)}\br\s
\label{C21}\] \noindent and  three nontrivial four--point
correlative WF: $C_{_{22} (\1\2|\3\4)}$ and

\[
\begin{array}{c}
\di C_{_{04} (\1 \2 \3 \4)}=
              \frac1{4!}\langle \vp_{(x_1)} \vp_{(x_2)} \vp_{(x_3)} \vp_{(x_4)}
              \rangle\s
-G_{(\1\2)}G_{(\3\4)}
\\[3mm]\hspace*{44mm} \di
-G_{(\1\3)}G_{(\2\4)}-G_{(\1\4)}G_{(\2\3)}
\end{array}
\]
\[
\begin{array}{c}\di C_{_{40} (\1 \2| \3 \4)}=
              \frac1{4!}\frac1{(N-1)^2}
         \langle \chi_{a (x_1)} \chi_{a (x_2)} \chi_{b (x_3)} \chi_{b(x_4)}
         \rangle\s-\\[3mm]\di
-\frac1{N-1}\left[(N-1)D_{(\1\2)}D_{(\3\4)}
+D_{(\1\3)}D_{(\2\4)}+D_{(\1\4)}D_{(\2\3)}\right]
\end{array}\label{C40}
\]
Correlative functions of higher rank are defined in the same way.
Strictly speaking, written above Wightman functions are divergent
and should be redefined by procedures of regularization and
renormalization. \footnote{In discussed formalism these operations
replace quantum--field operators product smoothing, used in
axiomatic quantum--field theory.}. In this paper the scheme of
dimensional regularization is used. This technique lets retain
exact thermodynamic relations between equations of state,
functionals and observables after renormalization.

In terms of Wightman functions momentum--energy tensor
(\ref{TofEI}) looks like:

\begin{equation}
\begin{array}{c}
\di \langle T_\mu^\nu\rangle=\lim_{x\rightarrow
x_1}\left(\partial_{\mu\hspace*{0.5mm}
(x)}\partial^\nu_{(x_{1})}G_{(x
x_1)}+(N-1)\partial_{\mu\hspace*{0.5mm}
(x)}\partial^\nu_{(x_{1})}D_{(x x_1)}\right)-
\delta_\mu^\nu\biggl(\frac{\mu^2 v^2}{2}-\frac{\la
v^4}{N}\hspace*{1mm}+\\[3mm] +\di\frac{\la}{N}
\left[3G_{(\0\0)}G_{(\0\0)}+(N^2-1)D_{(\0\0)}D_{(\0\0)}+2(N-1)G_{(\0\0)}D_{(\0\0)}\right]+
\frac{\la}{N}\left[C_{_{04} (\0\0\0\0)}+\right.
\\[3mm]\di +
\left.(N-1)^2 C_{_{40} (\0\0\0\0)}+ 2 (N-1) C_{_{22} (\0\0\0\0)}
+2v^2 C_{_{03} (\0\0\0)}+2v^2 (N-1) C_{_{21}
(\0\0|\0)}\right]\biggr)
\end{array}
\label{TEI O(N)}
\end{equation}

\paragraph{Derivation of the chain equations.}
First of all equations for WF forming momentum--energy tensor
should be obtained.

\noindent Equation of state for vacuum expectation value is
obtained by averaging (\ref{vp}):

\beq
v\left[-\mu^2+\frac{4\la}{N}\left(v^2+3G_{(\0\0)}+C_{_{03}(\0\0\0)}+
(N-1)D_{(\0\0)}+(N-1)C_{_{21}(\0\0|\0)}\right)
\right]=0.\label{O(N) condensat} \eeq

Averaging of  (\ref{chi}) gives
\[
2v\bl \vp\chi_a\br+\bl\vp^2\chi_a\br+\bl \chi_b \chi_b \chi_a\br
=0,
\]
which is identity owing to mentioned above symmetry
$\chi_a\to-\chi_a$.

Averaging equations (\ref{vp}) and (\ref{chi}) multiplied by one
or several operators, one obtain equations of the Bogolubov's
chain. After symmetrization, which is required in account of
symmetrical definition of WF, commutators arise:

\[
\Delta\vp_{(\0\1)}\equiv\vp_{(\0)}\vp_{(\1)}-\vp_{(\1)}\vp_{(\0)},
\hspace*{5mm}\di
\Delta\chi_{(\0\1)}\equiv\frac1{N-1}\left(\chi_{a(\0)}\chi_{a(\1)}-\chi_{a(\1)}\chi_{a(\0)}\right).
\]

Equations  for two--point WF look like: \beq\begin{array}{c}
\di
\partial_\mu\partial^\mu_{x}G_{(\0\1)}+\left[-\mu^2+\frac{4\la}{N}\left(3v^2+3G_{(\0\0)}+(N-1)D_{(\0\0)}\right)
\right]G_{(\0\1)}+
\\[3mm]\di+
\frac{4\la}{N}\left[\left(C_{_{04}(\0\0\0\1)}+(N-1)C_{_{22}(\0\0|\0\1)}\right)+
v^2\left(3C_{_{03}(\0\0\1)}+\right.\right.
\left.\left.(N-1)C_{_{21}(\0\0|\1)}\right)\right]=0
\end{array}\label{Gequation1}\eeq
\beq\begin{array}{c}\di
\partial_\mu\partial^\mu_{x}D_{(\0\1)}+\left[-\mu^2+\frac{4\la}{N}\left(v^2+G_{(\0\0)}+(N+1)D_{(\0\0)}\right)
\right]D_{(\0\1)}+
\\[3mm]\di+
\frac{4\la}{N}\left[
(N-1)C_{_{40}(\0\0|\0\1)}+C_{_{22}(\0\1|\0\0)} + 2v^2
C_{_{21}(\0\1|\0)}\right]=0
\end{array}\label{Dequation1}\eeq
Owing to homogeneity, isotropy and stationarity of the system
under consideration $G_{(\0\1)}$ and $D_{(\0\1)}$ depend  only on
module of two arguments difference $|\0-\1|$; consequently, when
$\0=\1$ their value are determined only by internal system
parameters. Thus, emerge in equations (\ref{Gequation}),
(\ref{Dequation}), all equations for WF and equations for
commutators values
\begin{equation}
m_1^2\equiv-\mu^2+\frac{4\la}{N}\left(3v^2+3G_{(\0\0)}
+(N-1)D_{(\0\0)}\right), \label{O(N) m_1}
\end{equation}
\begin{equation}
 m_2^2\equiv-\mu^2+\frac{4\la}{N}\left(v^2+G_{(\0\0)}+(N+1)D_{(\0\0)}
\right), \label{O(N) m_2}
\end{equation}
depend on the system state but not the 4--coordinates. Hence these
values play role of  parameters when chain equations are solved.

Taking into account the designations introduced, equations for
two--point WF can be rewritten as follows: \beq\begin{array}{c}
\di \partial_\mu\partial^\mu_{x}G_{(\0\1)}+m_1^2\;G_{(\0\1)}+
\\[1mm]\di+
\frac{4\la}{N}\left[\left(C_{_{04}(\0\0\0\1)}+(N-1)C_{_{22}(\0\0|\0\1)}\right)+
v^2\left(3C_{_{03}(\0\0\1)}+\right.\right.
\left.\left.(N-1)C_{_{21}(\0\0|\1)}\right)\right]=0,
\end{array}\label{Gequation}\eeq
\beq\begin{array}{c}\di
\partial_\mu\partial^\mu_{x}D_{(\0\1)}+m_2^2\;D_{(\0\1)}+
\\[1mm]\di+
\frac{4\la}{N}\left[
(N-1)C_{_{40}(\0\0|\0\1)}+C_{_{22}(\0\1|\0\0)} + 2v^2
C_{_{21}(\0\1|\0)}\right]=0.
\end{array}\label{Dequation}\eeq
Equation for WF $C_{_{03}(\0\1\2)}$ looks like:
\begin{equation}
\begin{array}{c} \di
\partial_\mu\partial^\mu_x C_{_{03} (\0\1\2)}+
   m_1^2\;C_{_{03}
   (\0\1\2)}+
\\[3mm]  \di
+\frac{4\la}{N}\left[G_{(\0\1)}\left(3 C_{_{03} (\0\0\2)}+
(N-1)C_{_{21} (\0\0|\2)}\vphantom{2^2} \right) + G_{(\0\2)}\left(
3C_{_{03} (\0\0\1)}+\right.\right.
\\[3mm]  \di
\left.+(N-1)C_{_{21} (\0\0|\1)} \vphantom{2^2} \right)
+(N-1)C_{_{22} (\0\0|\1\2)} +3C_{_{04} (\0\0\1\2)} +C_{_{05}
(\0\0\0\1\2)}+
\\[3mm]  \di
\left.+(N-1)C_{_{23} (\0\0|\0\1\2)}\right]
 =-\frac{24\la}{N}\left[ G_{(\0\1)}G_{(\0\2)}
+\frac1{12} \Delta\vp_{(\0\1)}\Delta\vp_{(\0\2)}\right].
\end{array}
\label{C03equation}
\end{equation}

Since $C_{_{03} (\0\1\2)}$ is invariant under transformations
$\0\leftrightarrow\1$, $\0\leftrightarrow\2$ and
$\1\leftrightarrow\2$, solution for this WF is a sum of solution
of (\ref{C03equation}) and solutions obtained from that by
permutations $\0\leftrightarrow\1$ and $\0\leftrightarrow\2$.

For $C_{_{21}(\0\1|\2)}$, which depends on $\vp$ and $\chi_a$
quantum fields, one should derive two equations, which describe
contributions of $\vp$ and $\chi_a$ respectively.

The first one is derived from (\ref{chi}),
\begin{equation}
\begin{array}{c} \di
\partial_\mu\partial^\mu_x C_{_{21} (\0\1|\2)}+
   m_2^2\;C_{_{21}
   (\0\1|\2)}+
   \\[3mm]  \di
+\frac{4\la}{N}\left[2 C_{_{22} (\0\0|\1\2)} +C_{23
(\0\0|\0\1\2)}+ D_{(\0\1)}C_{_{03} (\0\0\2)} +2 G_{(\0\2)}C_{_{21}
(\0\1|\0)}+\right.
\\[3mm]  \di
\left.+(N-1) C_{_{41} (\0\0|\0\1|\2)}
     +(N+1) D_{(\0\1)}C_{_{21} (\0\0|\2)}\right]=-\frac{8\la}{N}
     D_{(\0\1)}G_{(\0\2)},
\end{array}
\label{C211equation}
\end{equation}
\noindent and the second one from (\ref{vp}),
\begin{equation}
\begin{array}{c} \di
\partial_\mu\partial^\mu_{x_2} C_{_{21} (\0\1|\2)}+
   m_1^2\;C_{_{21}
   (\0\1|\2)}+
\\[3mm]  \di \frac{4\la}{N}\left[
3 C_{_{22} (\0\1|\2\2)} + C_{_{23} (\0\1|\2\2\2)} +(N-1) C_{_{41}
(\0\1|\2\2|\2)}+\right.
\\[3mm]  \di
\left.+(N-1)C_{_{40}(\0\1|\2\2)} +2\la D_{(\1\2)}C_{_{21}
(\0\2|\2)} +2\la G_{(\0\2)}C_{_{21} (\1\2|\2)}\right]=
\\[3mm]  \di
=-\frac{8\la}{N}\left[D_{(\1\2)}D_{(\0\2)}+\frac1{12}
\Delta\chi_{(\2\0)} \Delta\chi_{(\2\1)}\right].
\end{array}
\label{C212equation}
\end{equation}
Since $C_{_{21} (\0\1|\2)}$ is invariant under transformation
$\0\leftrightarrow \1$ solution for this WF is a sum of solutions
of (\ref{C212equation}) and (\ref{C211equation}), and solution of
(\ref{C211equation}) after permutation $\0\leftrightarrow \1$. The
same assertions are valid for all the other WF.

Equation for $C_{_{04} (\0\1\2\3)}$ is derived from (\ref{vp}):
\begin{equation}
\begin{array}{c} \di
\hspace*{5mm} \partial_\mu\partial^\mu_x C_{_{04} (\0\1\2\3)}+
    m_1^2\;C_{_{04} (\0\1\2\3)}+
    \\[4mm] \di\hspace*{5mm}
    +\frac{4\la}{N}\left[ G_{(\0\1)}\left[3C_{_{04} (\0\0\2\3)}+
    (N-1)C_{_{22} (\0\0|\2\3)}\right]+ G_{(\0\2)}\left[
    3C_{_{04} (\0\0\3\1)}+\right.\right.
\\[4mm] \di\hspace*{5mm}
    \left. +(N-1)C_{_{22} (\0\0|\1\3)}\right]
    +G_{(\0\3)}\left[
    3C_{_{04} (\0\0\1\2)}+(N-1)C_{_{22} (\0\0|\1\2)}\right]+
\\[4mm] \di\hspace*{5mm}
    +6 v^2\left[G_{(\0\1)}C_{_{03} (\0\2\3)}+
    G_{(\0\2)}C_{_{03} (\0\1\3)}+G_{(\0\3)}C_{_{03} (\0\1\2)}\right]+
\\[4mm] \di\hspace*{5mm}
   + v^2\left[C_{_{03} (\0\1\2)}(3C_{_{03} (\0\0\3)}+(N-1)C_{_{21} (\0\0|\3)})+
    C_{_{03} (\0\1\3)}(3C_{_{03} (\0\0\2)}+\right.
\\[4mm] \di\hspace*{5mm}
    \left.+(N-1)C_{_{21} (\0\0|\2)})+
    C_{_{03} (\0\2\3)}(3C_{_{03} (\0\0\1)}+(N-1)C_{_{21} (\0\0|\1)})\right]+
     \\[4mm] \di\hspace*{5mm}
     v^2 \left[3C_{_{05} (\0\0\1\2\3)}
+(N-1)C_{_{23} (\0\0|\1\2\3)}\right]+
    \left[C_{_{06} (\0\0\0\1\2\3)}+\right.
\\[4mm] \di \hspace*{5mm}
\end{array}
\end{equation}
\[
\begin{array}{c}
+\left.(N-1)C_{_{24} (\0\0|\0\1\2\3)}\right]
   = -\frac{\la}{N}\left[24 G_{(\0\1)}G_{(\0\2)}G_{(\0\3)}\right.+
\\[4mm] \di \hspace*{5mm}+
\Delta\vp_{(\0\1)}\Delta\vp_{(\0\2)}
    G_{(\0\3)}
    + \Delta\vp_{(\0\2)} \Delta\vp_{(\0\3)} G_{(\0\1)}
    +\Delta\vp_{(\0\1)} \Delta\vp_{(\0\3)}
 \left.    G_{(\0\2)}\right]
\end{array}
\label{C04equation}
\]

Since the chain equations are quite bulky, a simple criterion of
it's propriety is very useful. For example WF $C_{_{04}
(\0\1\2\3)}$ is invariant of permutation of it's arguments, so
that equation (\ref{C04equation}) should be invariant under
transformations $\1\leftrightarrow\2$, $\1\leftrightarrow\3$,
$\2\leftrightarrow\3$ (but, due to differential operator
$\partial_\mu\partial^\mu_x$, not the permutations involve $\0$).
Easy to check that the equation, written above, satisfy this
criterion.

Equation for WF $C_{_{40} (\0\1\2\3)}$ is derived from
(\ref{chi}):
\[
\begin{array}{c} \di
\hspace*{0mm}\partial_\mu\partial^\mu_x C_{_{40} (\0\1|\2\3)}+
    m_2^2\;
    C_{_{40} (\0\1\2\3)}+
\\[4mm] \di\hspace*{0mm}
    +\frac{4\la}{N}\left[D_{(\0\1)}((N+1)C_{_{40} (\0\0\2\3)} +
    C_{_{22} (\2\3|\0\0)} + 2v^2C_{_{21} (\1\3|\0)})+
D_{(\0\2)}(C_{_{40} (\0\0|\1\3)}+
    \right.
\\[4mm] \di\hspace*{0mm}
    + 2C_{_{40} (\0\1|\0\3)}
    + \frac1{N-1}C_{_{22} (\1\3|\0\0)} + \frac{2v^2}{N-1}C_{_{21}
    (\1\3|\0)})+D_{(\0\3)}(C_{_{40} (\0\0|\1\2)}+
\\[4mm] \di\hspace*{0mm}
    +2C_{_{40} (\0\1|\0\2)}
     +\frac1{N-1} C_{_{22} (\1\2|\0\0)} +
     \frac{2v^2}{N-1}C_{_{21} (\1\2|\0)})+ \frac{2v^2}{N-1}\left[
 C_{_{21} (\0\2|\0)}C_{_{21} (\1\3|\0)}+
     \right.
\\[4mm]\di\hspace*{0mm}
\left.   +   C_{_{21} (\0\3|\0)}C_{_{21} (\1\2|\0)} +(N-1)C_{_{21}
(\0\1|\0)}C_{_{21} (\2\3|\0)} \right]+    2 v^2 C_{_{41}
(\0\1\2\3|\0)}+
\end{array}
\]
\[
\begin{array}{c}
 \di
+\left. C_{_{42} (\0\1\2\3|\0\0)}+(N-1) C_{_{60}
(\0\0\0\1\2\3)}\right]=-\frac{8\la(N+1)}{N\cdot(N-1)}D_{(\0\3)}D_{(\0\2)}D_{(\0\1)}-
\\[4mm]\di-\frac{\la}{N(N-1)}\left[
\frac{(5N-7)}{3}\Delta\chi_{(\0\2)}\Delta\chi_{(\0\3)}D_{(\0\1)}
+\Delta\chi_{(\0\1)}\Delta\chi_{(\0\2)}D_{(\0\3)}+\right.
\\[4mm] \di\hspace*{85mm}
+ \Delta\chi_{(\0\1)}\Delta\chi_{(\0\3)}D_{(\0\2)}\biggr]
\end{array}
\label{C40equation}
\]

For WF $C_{_{22} (\2\3|\0\1)}$ one should obtain two independent
equations. The first one is derived from (\ref{vp}) by multiplying
by $\chi_ a(\2)\chi_a(\3)\vp(\1)$ and
$\chi_a(\3)\chi_a(\2)\vp(\1)$:
\[
\begin{array}{c} \di
\partial_\mu\partial^\mu_x C_{_{22} (\2\3|\0\1)}+
    m_1^2\;C_{_{22}
    (\2\3|\0\1)}+
\\[4mm] \di
   \frac{4\la}{N}\left[  G_{(\0\1)}\left((N-1)C_{_{40} (\0\0\2\3)}+3C_{_{22}
   (\2\3|\0\0)}\right)\right.
    +2\left[D_{(\0\2)}C_{_{22} (\0\3|\0\1)}+\right.
\\[4mm] \di
    \left.+D_{(\0\3)}C_{_{22} (\0\2|\0\1)}\right]
    +2\ v^2\left[3G_{(\0\1)}C_{_{21} (\2\3|\0)}+D_{(\0\3)}C_{_{21}
    (\0\2|\1)}+D_{(\0\2)}C_{_{21} (\0\3|\1)}\right]
\\[4mm] \di
    +3 v^2 C_{_{21} (\2\3|\0)}C_{_{03} (\0\0\1)}+
     v^2\left[(N-1)C_{_{21} (\2\3|\0)}C_{_{21} (\0\0|\1)}
+2C_{_{21} (\0\3|\1)}C_{_{21} (\0\2|\0)}+
     \right.
\\[4mm] \di
\left.+2C_{_{21} (\0\2|\1)}C_{_{21} (\0\3|\0)}\right]+ 3
v^2C_{_{23} (\2\3|\0\0\1)}+
    C_{_{24} (\2\3|\0\0\0\1)}+(N-1) v^2 C_{_{41} (\0\0\2\3|\1)}+
\\[4mm] \di
\left. +
    (N-1) C_{_{42} (\0\0\2\3|\0\1)}\right]
    =-\frac{8\la}{N}\left[D_{(\0\2)}D_{(\0\3)}G_{(\0\1)}+\frac1{12}
    \Delta\chi_{(\0\2)}\Delta\chi_{(\0\3)}G_{(\0\1)}\right],
\end{array}
\label{C221equation}
\]
and the second one from (\ref{chi}) by transformation
$\0\rightarrow\2$ and multiplying by $\chi(\3)\vp(\0)\vp(\1)$ and
$\chi(\3)\vp(\1)\vp(\0)$:
\[
\begin{array}{c} \di
\partial_\mu\partial^\mu_{x_2} C_{_{22} (\2\3|\0\1)}+
    m_2^2\;C_{_{22} (\2\3|\0\1)}+
\\[1mm] \di
    +\frac{4\la}{N}\biggl[ D_{(\2\3)}\left[C_{_{04} (\0\1\2\2)}+(N+1)C_{_{22} (\2\3|\0\0)}\right]+
        2\left[G_{(\0\2)}C_{_{22} (\2\3|\1\2)}\right.+
\end{array}
\]
\[
\begin{array}{c}
    \left.+G_{(\1\2)}C_{_{22} (\2\3|\0\2)}\right]
    +2 v^2\left[D_{(\2\3)}C_{_{03} (\0\1\2)}+G_{(\1\2)}C_{_{21}
    (\2\3|\0)}+\right.
\\[4mm]\di
\di\left. +G_{(\0\2)}C_{_{21} (\2\3|\1)}\right]    +
v^2\left[2C_{_{21} (\2\3|\2)}C_{_{03} (\0\1\2)}+C_{_{21}
(\2\3|\1)}C_{_{03} (\0\2\2)}+ \right.
\\[4mm] \di
+\left.    C_{_{21} (\2\3|\0)}C_{_{03} (\1\2\2)}\right]+
     v^2(N+1)\left[C_{_{21} (\2\2|\1)}C_{_{21} (\2\3|\1)}+
    C_{_{21} (\2\2|\1)}C_{_{21} (\2\3|\0)}\right]+
\\[3mm] \di
+ (N-1) C_{_{42} (\2\2\2\3|\0\1)}
    + C_{_{24} (\2\3|\0\1\2\2)}+
    2 v^2C_{23 (\2\3|\0\1\2)}\biggr]=
\\[4mm] \di
    =-\frac{8\la}{N}\left[D_{(\2\3)}G_{(\0\2)}G_{(\1\2)}+
    \frac1{12}\Delta\vp_{(\2\0)}\Delta\vp_{(\2\1)}D_{(\2\3)}\right]
\end{array}
\label{C222equation}
\]
Equations for commutators, which present almost in all chain
equations, look like: \beq
\partial_\mu\partial^\mu_{x}\Delta\vp_{(\0\1)}+m_1^2\;\Delta\vp_{(\0\1)}=0,
\label{comm_phi}\eeq \beq
\partial_\mu\partial^\mu_{x}\Delta\chi_{(\0\1)}+m_2^2\;\Delta\chi_{(\0\1)}=0.\label{comm_chi}
\eeq In high symmetry phase, where vacuum expectation value is
equal to zero, symmetry with regard to $\vp\rightarrow -\vp$
transformation is restored, consequently all WF which depend on
odd number of $\vp$ quantum--field operators are identically equal
to zero. The chain equations for this phase are easily derived
from those for low--symmetry phase.

\paragraph{The chain reduction.}
Two approaches to the chain reduction are known:
model--approximative and iterative.

In model--approximative approach a few first chain equations are
solved, and the higher rank Wightman functions are evaluated as a
certain combination of lower--rank WF according to an algorithm
specified. Use of this approach in the considered formalism is
difficult due to necessity of WF redefinition: in order to
renormalize divergent integrals one should know energy spectrum,
which, in turn, can be obtained only after renormalization.

In iterative approach corrections to a basis approximation are
calculated. Choice of basis approximation is dictating by the
chain mathematical structure.

A chain equation is a generalized D'Alamber equation
\[
\partial_\mu\partial^\mu_{x} C_{_{nm} (\0\ldots
x_m)}+m^2\;C_{_{nm} (\0\ldots x_m)}={\rm f}\;(\0\ldots x_m)
\]
with combinations of correlative functions and field commutators
as sources, which solution is a sum of general solution of the
corresponding homogeneous equation and partial solution concerned
with the sources.

Solution of homogeneous equation, considered in chapter \ref{PSP},
should be taken into account only for two--point WF $G_{(\1\2)}$
and $D_{(\1\2)}$, which form effective masses $m_1$, $m_2$. For
correlative functions, which describe weak effects of
many--particles interaction, in considered iterative approach one
should take into account only solutions concerned with the
sources. Since sources contain small constant of
self--interaction, correlative WF $C_{_{03} (\1\2\3)}$, $C_{_{21}
(\1\2|\3)}$ are of the first order of vanishing in compare with
$G_{(\1\2)}$ and $D_{(\1\2)}$.

Independently of choice of basis approximation the following
algorithm of calculations can be formulated:

I.1. Derive solutions of the chain equations (considered in a
certain approximation) keeping $m_1$, $m_2$ and $v$ as parameters.
As a result one obtain WF as functions of coordinates, effective
masses, order parameter and temperature:

\[
W_{(\0\1\ldots x_n x_m)}=W(\0\1\ldots x_n x_m,m_1,m_2,v,T)
\]

I.2. Calculate WF at coincident points.  Owing to homogeneity,
isotropy and stationarity of the system under consideration, the
result depends on temperature, $m_1$, $m_2$ and $v$, but not
4--coordinates:

\[
W_{(\0\0\ldots\0\0)}=W(m_1,m_2,v,T)
\]
I.3. Substitute Wightman functions, calculated at coincident
points, in (\ref{O(N) m_1}), (\ref{O(N) m_2}) and (\ref{O(N)
condensat}) and solve this system of nonlinear equations for $v$,
$m_1$, $m_2$. After substitution of solutions
\[
m_1=m_1(T),m_2=m_2(T),v=v(T)
\]
one obtain WF as functions of coordinates and temperature.

I.4. By known dependence WF on temperature calculate thermodynamic
observables using momentum--energy tensor (\ref{TEI O(N)}).

The chain reduction basis approximation corresponds to allowing
only for two--point $G_{(\0\1)}$ and $D_{(\0\1)}$ Wightman
functions in equations (\ref{Gequation}) and (\ref{Dequation}),
equation of state for order parameter (\ref{O(N) condensat}) and
momentum--energy tensor (\ref{TEI O(N)}).

When calculating corrections to the basis approximation, which
assumes calculation of higher rank  WF, already known two--point
Wightman functions are used:

II.1. Substitute $G_{_0 (\1\2)}$ and $D_{_0 (\1\2)}$, calculated
in basis approximation, in equations for correlative WF, in
particular $C_{_{03} (\1\2\3)}$ and $C_{_{21} (\1\2|\3)}$, and
obtain partial solution, concerned with sources.

II.2. Substitute obtained solutions for correlative WF into the
chain equations for $G_{(\1\2)}$, $D_{(\1\2)}$, obtain their
solutions and so on...

II.3. Calculate $v$, $m_1$, $m_2$ and thermodynamic observables on
temperature dependence (according to points I.2 --- I.4 of
algorithm).

It is easy to  see, that corrections are of $O(N^{-1})$ order of
vanishing, thus at the limit $N\rightarrow\infty$ the basis
approximation becomes exact solution of the chain equations.

\section{Hartree--Fock approximation.\label{PSP}}
The Bogolubov's chain basis reduction approximation is called
Hartree--Fock approximation (HFA)\footnote{Which is often
mentioned as ``\hspace*{0.1mm}mean field approximation'',
``\hspace*{0.1mm}Hartree approximation''.}

\paragraph{The chain equations in Hartree--Fock approximation.}
According to algorithm, stated above, at first step equations for
two--point functions, with $m_1$ and $m_2$ considered as
parameters, are written out:
\[
\begin{array}{c}
\partial_\mu\partial^\mu_x G_{_0 (\0\1)}+m_1^2\; G_{_0(\0\1)}=0
,\partial_\mu\partial^\mu_x  D_{_0(\0\1)}+ m_2^2\; D_{_0(\1\0)}=0.
\label{PSPforD&G}
\end{array}
\]
These equations are similar to those for noninteracting fields,
thus it is natural to represent $\vp$ and $\chi_a$ as follows:
\begin{equation}
\begin{array}{c}
\di \vp=\sum\limits_\vecp\frac1{\sqrt{2\ve_{_1 \vecp}}}(a_\vecp
e^{-i\ve_{_1 \vecp} t}+ a_{-\vecp}^+e^{i\ve_{_1 \vecp}
t})e^{i\vecp \vecx},
              \qquad \ve_{_1 \vecp}^2=\vecp\,{}^2+m_1^2,
\\[5mm] \di
\chi_a=\sum\limits_\vecp\frac1{\sqrt{2\ve_{_2
\vecp}}}(b_{p(a)}e^{-i\ve_{_2 \vecp} t}+ b_{-\vecp
(a)}^+e^{i\ve_{_2 \vecp} t})e^{i\vecp \vecx},
               \qquad \ve_{_2 \vecp}^2=\vecp\,{}^2+m_2^2,
\end{array}
\label{field structure}
\end{equation}
with usual commutators for $a_\vecp, b_{\vecp (a)}, a^+_\vecp,
b^+_{\vecp (a)}$ operators. This representation corresponds to
description of the system as an ideal gas of particles with masses
dependent on temperature.

At the second step two--point WF are calculated in coincident
points:
\begin{equation}
\begin{array}{c}
\di \la G_{_0(\0\0)}=\la \sum_\vecp \frac{N_{_1 \vecp}+1/2}{V
\ve_{_1 \vecp}} =\la \sum\limits_\vecp\frac{1}{V \ve_{_1
\vecp}}\cdot\frac{1}{\exp\frac{\ve_{_1 \vecp}}{T}-1} +
\frac12\left(\la\sum\limits_\vecp\frac{1}{V \ve_{_1
\vecp}}\right)_{ren},
\\[6mm]
\di \la D_{_0(\0\0)}=\la \sum\limits_{\vecp}\frac{N_{_2
\vecp}+1/2}{V \ve_{_2 \vecp}} =\la \sum\limits_{\vecp}\frac{1}{V
\ve_{_2 \vecp}}\cdot\frac{1}{\exp\frac{\ve_{_2 \vecp}}{T}-1} +
\frac12\left(\la\sum\limits_\vecp\frac{1}{V \ve_{_2
\vecp}}\right)_{ren}. \label{gf}
\end{array}
\end{equation}
\noindent As it was already mentioned, divergent terms were
renormalized in the framework of dimensional regularization
method. Essential part of the method is that a theory is initially
formulated in non-integer dimension space and transmutation
parameters present in Lagrangian as factor multiplying
dimensionless constants of interaction. Indeed, in Lagrangian
(\ref{lagrangian}), written for space--time of dimension
$D=1+n=1+3-2\ve$, number of dimension of self--interaction
constant $\la$ is $L^{n-3}$; dimensional transmutation parameter
is defined as $\la_n\!=\!\la l^{n-3}$. Analytic continuation to
integer dimension space $\ve\to 0$, $D\to 1+3$ is made only after
subtraction of $\ve^{-1}$ poles (this operation is equivalent to
redefinition of inoculating constants of the model). After
renormalization of divergent terms (\ref{gf}) one obtain:
\[
\biggl(\la_n\sum\limits_{\vecp}\frac{1}{2 V
\ve_{\vecp}}\biggr)_{ren}=
\la\frac{m^2}{16\pi^2}\ln\biggl(\frac{m^2}{\Lambda^2}\biggr)
,\hspace*{5mm} \Lambda^2=\frac{4\pi l^{-2}}{e^{{\cal C}-1}},
\label{renorm1}
\]
where ${\cal C}=0.5772157$ is Euler constant.

\noindent At third step renormalized WF are substituted to
(\ref{O(N) m_1}), (\ref{O(N) m_2}) and (\ref{O(N) condensat}).

It should be noted, that independently of renormalization method
used, equation of state for order parameter (\ref{O(N) condensat})
in Hartree--Fock approximation is written as:

\beq v\left[m_1^2-\frac{8\la}{N}v^2\right]=0 \label{order_par_eq}
\eeq \noindent Gap equations for effective masses (\ref{O(N) m_1})
and (\ref{O(N) m_2}) for further purposes should be rewritten as
follows:
\begin{equation}
\begin{array}{c}\di
\left[J_1(m_1,T)+\frac{m_1^2}{16\pi^2}\ln\left(\frac{m_1^2}{\Lambda^2}\right)\right]-
\frac{N}{8\la(N+2)}\left[(N+1)m_1^2-(N-1)m_2^2+2\mu^2\right]
+v^2=0,
\\[6mm]\di
(N-1)\left[J_1(m_2,T)+\frac{m_2^2}{16\pi^2}\ln\left(\frac{m_2^2}{\Lambda^2}\right)\right]-
\frac{N(N-1)}{8\la(N+2)}\left[3m_2^2-m_1^2+2\mu^2\right]=0,
\end{array}
\label{urav_psp_renorm}
\end{equation}
where following designations for integrals over Bose--Einstein
distribution are introduced:
\[
J_n(m,T)=\frac1{2\pi^2}\int_0^{\infty}\frac{p^{2n} d p
}{\sqrt{p^2+m^2}} \cdot\frac{1}{\exp\frac{\sqrt{p^2+m^2}}{T}-1},
\qquad n=0,1,2, \label{temp_integral}
\]

\noindent Temperature integrals satisfy recurrent relations:
\[
\di \frac{\partial J_n(m,T)}{\partial m}=-(2n-1)m
J_{n-1}(m,T),
\hspace*{2mm} \frac{\partial J_n(m,T)}{\partial
T}=\frac{2n}{T}J_n(m,T)+(2n-1)\frac{m^2}{T}J_{n-1}(m,T)
\label{recurent}
\]
Gap equation for effective mass at high--symmetry phase is
obtained from (\ref{urav_psp_renorm}) by zero filling of order
parameter. It is easy to see, that at high--symmetry phase
$m_1=m_2\equiv m$.
\\[1mm]
It follows from (\ref{order_par_eq}) and (\ref{urav_psp_renorm})
that in Hartree--Fock approximation the following effects are
taken into account:

1)vacuum influence on quasi--particles properties (order parameter
$v$ in definition of effective masses).

2)whole particles influence on condensate and properties of each
quasi--particle (temperature integrals in equations for condensate
and effective masses).

3)zero-point oscillations of quantum fields influence on
condensate and properties of each quasi--particle (renormalized
vacuum integrals in equations for $v$ and effective masses).

\paragraph{Generating functional.}
Lagrangian (\ref{lagrangian}) is a ``generating functional'' for
operator equations of state, cause it reproduce these equations on
it's extremals. Since Bogolubov's chain is equivalent to set of
operator equations, a ``generating functional'' which reproduce
chain equations must exist. The same assertion is valid for the
chain reduced in any approximation.

In self--consistent approximation this functional should reproduce
gap equations for effective masses and equation of state for order
parameter, or their linear combination.

By equations (\ref{order_par_eq}) and (\ref{urav_psp_renorm})
generating functional, which satisfy requirements
\begin{equation} \left(\frac{\partial F}{\partial
m_1}\right)_{m_2,\hspace*{0.5mm}v}=0,\hspace*{2mm}
\left(\frac{\partial F}{\partial m_2}\right)_{m_1,\hspace*{0.5mm}
v}=0,\hspace*{2mm} \left(\frac{\partial F}{\partial
v}\right)_{m_1,\hspace*{0.5mm}m_2}=0. \label{GenFuncDiff}
\end{equation}
is reconstructed up to an arbitrary constant. It's explicit form:
\begin{equation}
\begin{array}{c}\di
F(T,m_1,m_2,v)=-\frac13\left[J_2(m_1,T)+(N-1)J_2(m_2,T)\right]+U(m_1,m_2,v),
\\[4mm]\di U(m_1,m_2,v)=\frac{m_1^4}{64\pi^2}\ln\left(\frac{m_1^2}{\sqrt{e}\Lambda^2}\right)+
(N-1)\frac{m_2^4}{64\pi^2}\ln\left(\frac{m_2^2}{\sqrt{e}\Lambda^2}\right)-
\\[6mm]
\di -\frac{N}{8\la (N+2)}\left[\di
\frac{(N+1)m_1^4+3(N-1)m_2^4}{4}- \frac{N-1}{2}\hspace*{1mm} m_1^2
\hspace*{0.5mm} m_2^2+\right.
\\[4mm]
\end{array}
\label{GenFunc}
\end{equation}
\[
\begin{array}{c}\di
+\mu^2(m_1^2+(N-1)m_2^2)\biggr] +\di\frac{m_1^2 v^2}{2}-\frac{2\la
v^4}{N}-\frac{N^2\mu^4}{16\la(N+2)}.
\end{array}
\]

\paragraph{Gibbs potential.}
The goal of the theory of phase transitions is a calculation of
transition temperature, phase co-existence region, and
thermodynamic observables. The last can be obtained from
momentum--energy tensor components (\ref{TEI O(N)}), which for
homogeneous and isotropic system is diagonal $\langle
T^\nu_\mu\rangle=\rm{diag}(\ve,-p,-p,-p)$. Since state of the
system under consideration is defined by temperature and volume,
one should use Gibbs potential for thermodynamic description. Due
to chemical potential equality to zero, potential of free energy
coincides with $\Om$-potential, which is easily calculated via
momentum--energy tensor spatial components:
\begin{equation}
\Omega=-\lim_{n\rightarrow 3}p_nV_n=\lim_{n\rightarrow
3}\frac{\langle T^i _i \rangle V_n}{n}. \label{omega-pot}
\end{equation}
As (\ref{omega-pot}) is divergent, it should be renormalized. In
(\ref{TEI O(N)}) two--point WF \hspace*{1mm} $G_{(\0\0)}$ and
$D_{(\0\0)}$ are expressed through effective masses and,
consequently, are not divergent. ``Kinetic'' momentum--energy
tensor terms
\[
\lim_{n\rightarrow 3}
\frac{V_n}{n}\cdot\left(\langle\partial_i\vp\hspace*{1mm}\partial^i\vp\rangle+
\langle\partial_i\chi_a\hspace*{1mm}\partial^i\chi_a\rangle\right)
\]
except directly dependent on temperature items,
\[
-\frac13\sum_\vecp \frac{\vecp^2 N_{_1\vecp}}{\ve_{_1
\vecp}}-\frac{N-1}3\sum_\vecp \frac{\vecp^2 N_{_2\vecp}}{\ve_{_2
\vecp}}
\]
contain divergent vacuum terms,
\[
\di\lim_{n\rightarrow 3} \frac{V_n}{n}\cdot\left(-\sum_\vecp
\frac{\vecp^2 }{2 V_n \ve_{_1 \vecp}}-(N-1)\sum_\vecp
\frac{\vecp^2 }{2 V_n\ve_{_2 \vecp}}\right)
\]
\noindent which should be renormalized. The parameter of
dimensional regularization method is introduced via volume $V_n$
in non--integer dimension space, which is connected with
integer--dimension space volume by simple relation:
\[
V_n=N l^n=N l^3\cdot l^{-2\ve}=V l^{-2\ve}.
\]
Renormalization of divergent terms leads to the following result:
\[
\lim_{n\rightarrow 3} \frac{V_n}{n}\cdot\left(-\sum_\vecp
\frac{\vecp^2 }{2 V_n
\ve_{\vecp}}\right)=V\cdot\frac{m^4}{64\pi^2}\ln\left(\frac{m^2}{\sqrt{e}\Lambda^2}\right)
\]
Substitution of results of renormalization of divergent terms and
two--point WF, expressed via effective masses and order parameter,
to (\ref{TEI O(N)}) gives:
\begin{equation}
\di \frac{{\cal F}(T,V)}{V}=-\frac13[J_2(T,m_1(T))+
J_2(T,m_2(T))]+U(m_1(T),m_2(T),v(T)) \label{Free_energy}
\end{equation}
Free energy density (\ref{Free_energy}) is distinguished into two
parts: the first one describes contribution of ideal gas
(temperature integrals over Bose--Einstein distribution), and the
second one contribution of so called ``self--consistent field''
(indirectly dependent on temperature).

It follows from (\ref{GenFunc}) and (\ref{Free_energy}), that
generating functional, being considered on solution of gap
equations for effective masses and equation of state for order
parameter, turns into equilibrium free energy density.
\begin{equation}
F(T,m_1(T),m_2(T),v(T))=\frac{{\cal F}(T,V)}{V}
\label{GenFuncProperty}
\end{equation}
For this reason arbitrary constant in (\ref{GenFunc}) was chosen
to be equal to $\di-\frac{N^2\mu^4}{16\la(N+2)}$.

\paragraph{Thermodynamic observables.}
Specific heat capacity, entropy and sonic speed are obtained from
equilibrium free energy by usual thermodynamic relations.
\[
\begin{array}{c}
\di s=\frac{S}{V}=-\frac1{V}\left(\frac{\partial {\cal
F}}{\partial T}\right)_{V} \di,
c_V=\frac{C_V}{V}=-\frac{T}{V}\left(\frac{\partial^2 {\cal F}
(T)}{\partial T^2}\right)_{V} \di, u^2=\frac{s}{c_V}.
\end{array}
\]
Due to mentioned property of generating functional
(\ref{GenFuncProperty}) and in view of (\ref{GenFuncDiff}) are
identities on solution of gap equations and equation of state
\[
\begin{array}{c}\hspace*{10mm}\di s=-\frac1{V}\left(\frac{\partial {\cal
F}(T,V)}{\partial T}\right)_V=-\left(\frac{\partial F}{\partial
T}\right)_{v,m_1,m_2},\hspace*{2mm} \di
c_V=-\frac{T}{V}\left(\frac{\partial^2 {\cal F}(T,V)}{\partial
T^2}\right)_V=
\\[5mm]\di
=-T\left[\left(\frac{\partial^2 F}{\partial
T^2}\right)_{v,m_1,m_2}+\left(\frac{\partial^2 F}{\partial T
\partial v}\right)_{m_1,m_2}\frac{d v}{d T}\right.
\left.+\left(\frac{\partial^2 F}{\partial m_1\partial T
}\right)_{v,m_2}\frac{d m_1}{d T} +\left(\frac{\partial^2
F}{\partial T \partial m_2}\right)_{v,m_1}\frac{d m_2}{d
T}\right].
\end{array}
\]
Effective masses and condensate order parameter derivatives are
obtained by derivation of (\ref{order_par_eq}) and
(\ref{urav_psp_renorm}), considered as identities. Result of
calculation is given below. Specific heat capacity in
low--symmetry phase:
\[
\begin{array}{c}\di
c_V=\frac1{T}\biggl(4\left[J_2(m_1,T)+(N-1)J_2(m_2,T)\right]+
5\left[m_1^2J_1(m_1,T)+(N-1)m_2^2 J_1(m_2,T)\right] +
\\[5mm]\di
+\left[m_1^4 J_0(m_1,T)+(N-1)m_2^4J_0(m_2,T)\right]+
\frac{4\la(N+2)}{\delta\cdot N}\left[
f_2^2(m_1,T)\left[3-\di\frac{4\la(N+2)}{N}f_1(m_2,T)\right]-\right.
\\[4mm]\di\left.
-f_2^2(m_2,T)(N-1)\left[\di\frac{4\la(N+2)}{N}f_1(m_1,T)+1\right]+
2(N-1)f_2(m_1,T)f_2(m_2,T)\right]\biggr).
\end{array}
\label{LSPheatcapacity}
\]
Here following designations are introduced:
\[
f_1(m,T)\equiv\A,\hspace*{3mm} f_2(m,T)\equiv2 J_1(m,T)+ m^2
J_0(m,T),
\]
\[
\delta\equiv{N-1-\left(\di\frac{4\la(N+2)}{N}f_1(m_2,T)-3\right)
\left(\di\frac{4\la(N+2)}{N}f_1(m_1,T)+1\right)}. \label{delta}
\]
Specific heat capacity in high--symmetry phase:
\[
c_V=\frac{N}{T}\left(4 J_2(m,T)+5 m^2 J_1(m,T)+m^4
J_0(m,T)-\di\frac{f_2^2(m,T)}{f_1(m,T)-\di\frac{N}{2\la (N+2)}}
\right).
\]
Specific entropy in high--symmetry phase is easily obtained from
that in low--symmetry phase:
\[
s=\frac1{T}\left(\frac43\left[J_2(m_1,T)+(N-1)J_2(m_2,T)\right]+
\left[m_1^2 J_1(m_1,T)+(N-1)J_1(m_2,T)\right]\right)
\]

\paragraph{Phases stability conditions.}
Free energy density (\ref{Free_energy}), obtained earlier by
clearly thermodynamic relation (\ref{omega-pot}), can be
calculated via statistical sum:
\begin{equation}
{\cal F}=-T\ln Z. \label{FreeEnStatSum}
\end{equation}
The system energy spectrum, which is needed for the calculation,
is determined by infill numbers, effective masses and condensate
values:
\[
E=\sum_\vecp \ve_{1 \vecp}\hspace*{0.5mm} N_{1
\vecp}+(N-1)\sum_\veck \ve_{2 \veck}\hspace*{0.5mm} N_{2
\veck}+V\cdot U(m_1,m_2,v),
\]
where $N_{1 \vecp}$ and $N_{2 \vecp}$ are numbers of $\vp$ and
$\chi_a$ particles respectively. For the theory to be
self--consistent in approximation used, one should sum over infill
numbers $N_{1 \vecp}$ and $N_{2 \vecp}$, but not the effective
masses and order parameter, considering $m_1$, $m_2$ and $v$ as
constants (as it was stated in \cite[\S 30]{kirznits}). After
summation statistical sum is represented as a product of two
terms: the first one corresponds to ideal gas and the second one
to ``self--consistent'' field. Taking the logarithm and summing
over impulses gives a function which coincide with generating
functional. Consideration of gap equations for effective masses
(\ref{urav_psp_renorm}) and equation of state for order parameter
(\ref{order_par_eq}) leads to the specific equilibrium free energy
density (\ref{Free_energy}).

Keeping symmetrical parameters as constants, when calculating
statistical sum, corresponds to the algorithm of calculation of
non--equilibrium Landau statistical sum $Z_L$ \cite{landau}, which
allows to obtain non-equilibrium free energy functional ${\cal
F}_L$. Since the only symmetrical parameter in our case is
condensate $v$, to obtain non--equilibrium Landau functional, one
should substitute effective masses on order parameter dependence
into generating functional.

The functional obtained $F(T,v)=F(T,m_1(T,v),m_2(T,v),v)$
reproduce equation of state for order parameter:
\[
\frac{d}{d v}F(T,v)=\left(\frac{\partial F}{\partial
m_1}\right)_{T,m_2,v}\frac{d m_1}{d v}+ \left(\frac{\partial
F}{\partial m_2}\right)_{T,m_1,v}\frac{d m_2}{d
v}+\left(\frac{\partial F}{\partial
v}\right)_{T,m_1,m_2}=\left(\frac{\partial F}{\partial
v}\right)_{T,m_1,m_2},
\]
and on solution of that equation $v=v(T)$ it turns to free energy
density, i.e. it possess all properties of non--equilibrium
functional.

A phase stability condition is order parameter Landau functional
second derivation positivity:
\begin{equation}
\frac{d^2 F}{d v^2}=m_1^2-\frac{24\la}{N}v^2+2\hspace*{1mm}
m_1\hspace*{0.5mm} v\frac{d m_1}{d v}. \label{d2Fdv2}
\end{equation}
Inclusion of formal external classical source, interacting with
condensate, in Lagrangian changes equation of state for
\[ {\cal L}\longrightarrow{\cal L}+\rho_a\phi_a,\hspace*{2mm}
\rho_a=\delta_{a,\hspace*{0.5mm} _N}\cdot\eta\] \noindent order
parameter, which now depends on external source $v=v(\eta)$. In
Hartree--Fock approximation:
\[
v\left[m_1^2-\frac{8\la}{N}v^2\right]=\eta.
\]
Generalized sensitivity at $\eta\rightarrow 0$ turns out to be in
inverse proportion with stability condition (\ref{d2Fdv2}):
\[\di\left(\frac{d v}{d \eta}\right)_{\eta\rightarrow
0}=\left(\frac{d^2 F}{d v^2}\right)^{-1}.\]

Being usual in Landau phase transitions theory, this result proves
correctness of non--equilibrium free energy and stability
condition (\ref{d2Fdv2}).

Appearing in (\ref{d2Fdv2}) condensate derivative of  mass is
obtained by order parameter differentiating of gap equations for
effective masses (\ref{urav_psp_renorm}), considered as
identities. In particular, low--symmetry phase stability condition
looks like:
\[
\di\frac{d^2 F}{d v^2}=\di2 m_1^2
\frac{N-1-\left(\di\frac{4\la(N+2)}{N}f_1(m_2,T)-3\right)
\left(\di\frac{4\la(N+2)}{N}f_1(m_1,T)+1\right)}
{\left(\di\frac{4\la(N+2)}{N}f_1(m_2,T)-3\right)
\left(\di\frac{4\la(N+2)}{N}f_1(m_1,T)-N-1\right)-N+1}.
\label{LSPd2Fdv2}
\]
In the vicinity of critical temperature $T_{c_2}$ stability
condition denominator is a slow variable function. Numerator
temperature derivative, which designation was introduced earlier,
$\di\delta^{\hspace*{0.5mm}'} _T\sim\delta^{-1}$, this implies
that $\delta\sim\sqrt{T_{c_2}-T}$ at the vicinity of $T_{c_2}$.

Specific heat capacity of low--symmetry phase contains $\delta$ in
inverse proportion, thus it is formally  divergent at $T_{c_2}$
temperature. Sonic speed, which is inverse proportion with heat
capacity, tends to zero.

Stability condition of high--symmetry phase proportional to
effective mass squared
\[
\frac{d^2F}{d v^2}=m^2,
\]
\noindent which allows to obtain value of critical temperature
$T_{c_1}$ analytically:
\[
\di T_{c_1}=\mu\sqrt{\frac{3 N}{\lambda (N+2)}}.
\]
\paragraph{Numerical calculations in Hartree--Fock approximation.}
Results of numerical calculations of effective masses, order
parameter and thermodynamical observables on temperature
dependence for $N=2$, $N=4$ and $N=9$ and $\la=0.01\cdot 2\pi^2$
are given below.

If parameter $\Lambda$ is chosen to be
$\Lambda=m_{1\hspace*{1mm}vac},\hspace*{1mm}
m_{1\hspace*{1mm}vac}\equiv m_1(T=0)$, gap equations for effective
masses and equation for order parameter posses a solution
$m_1^2=2\mu^2, \hspace*{1mm}m_2^2=0$, so that quanta of $\chi_a$
fields turn to Goldstone bosons.

It is convenient to normalize effective masses, order parameter
and observables to $m_{1\hspace*{1mm}vac}$. Graphs of their
temperature dependence are given in mentioned normalization ($\di
T\Rightarrow\frac{T}{m_{1 vac}}$ and so forth.).

\begin{center}
\includegraphics[width=16.5cm]{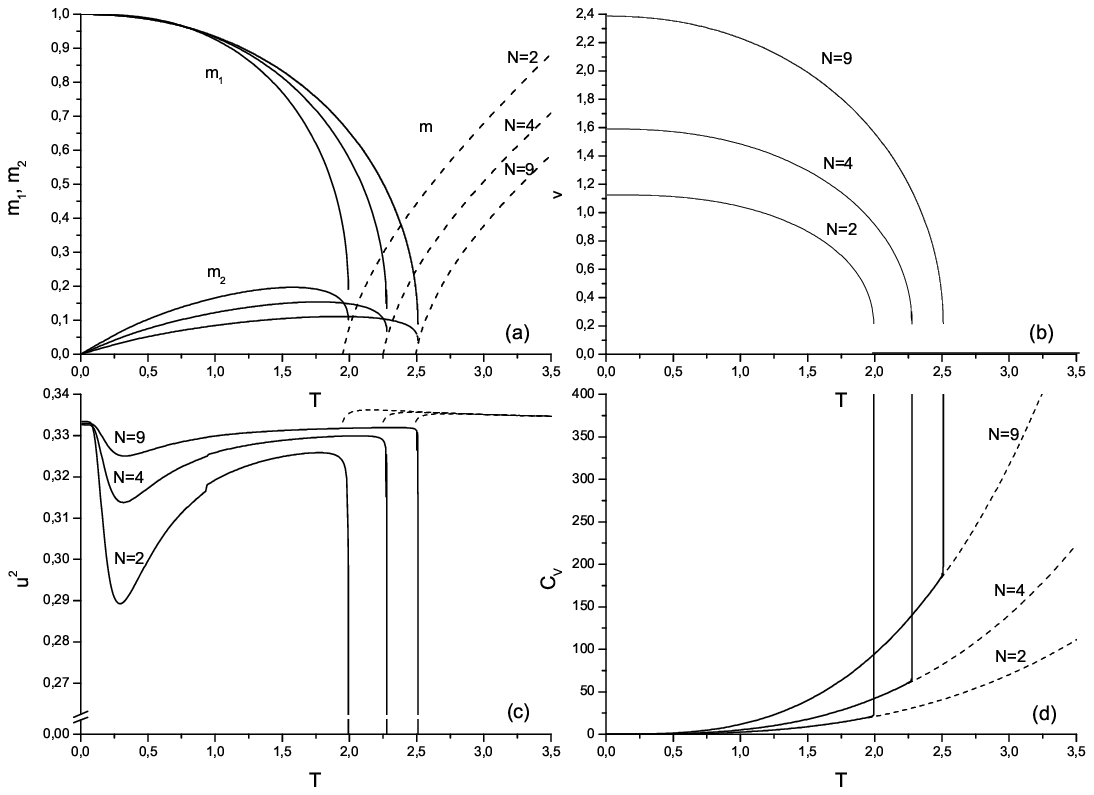}
\\
\rm Fig.1 Temperature dependence of effective masses, order
parameter, sonic speed and specific heat capacity in Hartree--Fock
approximation.
\end{center}

Equations (\ref{urav_psp_renorm}) and (\ref{order_par_eq}) in
low--symmetry phase  admit thermodynamically stable in temperature
region 0 .. $T_{c_2}$ (solid lines at figures 1.(a) and 1.(b)) and
in temperature region $T_{c_1}$ .. $T_{c_2}$ thermodynamically
unstable (not shown) branches of solutions. In high--symmetry
phase only one, stable, branch of solutions exists (dashed line).

Below temperature $T_{c_1}$ in high--symmetry phase and above
$T_{c_2}$ in low--symmetry phase, where stability conditions of
the phases turn to zero, equations for effective masses
(\ref{urav_psp_renorm})  and order parameter (\ref{order_par_eq})
admit no solutions. When calculating correction to Hartree--Fock
approximation, this criterion allows to easily find phases
stability thresholds.

As it follows from order parameter on temperature dependence, at
phase equilibrium point $T_{c_1}<T_c<T_{c_2}$, where equality of
free energies of both phases is achieved, order parameter is small
but not zero, consequently for finite $N$ first type (close to
second type) phase transition takes place. Decreasing of
condensate value at equilibrium point $T_c$ and decreasing of
phase coexistence region to $T_c$ ratio $\di\xi\equiv\frac{\Delta
T}{T_c}=\frac{T_{c_2}-T_{c_1}}{T_c}$ with increase $N$
$(\xi|_{N=2}\approx0.021$,  $\xi|_{N=4}\approx 0.011$,
$\xi|_{N=9}\approx0.006)$ indicates that at the limit
$N\rightarrow\infty$ phase transition of second type takes place.
Due to zero $\chi_a$ fields quanta masses, sonic speed is not zero
at zero temperature.

Numerical calculations confirm mentioned above conclusion of heat
capacity divergence at vicinity of $T_{c_2}$. However, it is well
known, that heavy fluctuations in this region ``blur'' jump of
heat capacity and other thermodynamic observables.

Corrections to Hartree--Fock approximation turn out large at
phases equilibrium point, which confirms inadaptability of the
approximation at this temperature and, consequently, formality of
made above conclusion about heat capacity divergence.

\section{Corrections to Hartree--Fock approximation.\label{Popravka}}
Calculation of corrections to Hartree--Fock approximation, which
implies calculation of higher--rank WF, is important not only for
estimation of the Hartree--Fock approximation temperature region
of adaptability, but also for proof of correctness of mentioned in
\ref{PSP} renormalization method.

\paragraph{Algorithm of calculations.}
Since equation for order parameter (\ref{O(N) condensat}), which
can be rewritten as follows
\begin{equation}
v\left[m_1^2-\frac{8\la v^2}{N}+\frac{4\la}{N}\left(
C_{_{03}(\0\0\0)}+(N-1)C_{_{21}(\0\0|\0)} \right) \right]=0,
\label{O(N)condensat_2}
\end{equation}
contains three--point WF, at first step one should derive partial
solution of equation for WF $C_{_{03} (\1\2\3)}$ and $C_{_{21}}$
concerned with sources. Proximate system of equations for
three-point WF is derived from (\ref{C03equation}),
(\ref{C211equation}) and (\ref{C212equation}) by neglecting of
higher--rank WF:
\begin{equation}
\begin{array}{c} \di
\partial_\mu\partial^\mu_x C_{_{03} (\0\1\2)}+
   m_1^2\; C_{_{03}
   (\0\1\2)} \approx-\frac{24\la}{N}\left[ G_{(\0\1)}G_{(\0\2)}
+\frac1{12} \Delta\vp_{(\0\1)}\Delta\vp_{(\0\2)}\right],
\label{C03eq}
\end{array}
\end{equation}
\begin{equation}
\begin{array}{c} \di
\partial_\mu\partial^\mu_{x_2} C_{_{21} (\0\1|\2)}+
   m_1^2\; C_{_{21}
   (\0\1|\2)}\approx-\frac{8\la}{N}\left[D_{(\2\1)}D_{(\2\0)}+\frac1{12}
\Delta\chi_{(\2\0)} \Delta\chi_{(\2\1)}\right], \label{C211eq}
\end{array}
\end{equation}
\begin{equation}
\begin{array}{c} \di
\partial_\mu\partial^\mu_x C_{_{21} (\0\1|\2)}+
   m_2^2\; C_{_{21}
   (\0\1|\2)}\approx-\frac{8\la}{N}
     D_{(\0\1)}G_{(\0\2)}.\label{C212eq}
\end{array} \end{equation}
In the framework of the iterative procedure used, for solving
equations (\ref{C03eq}) and (\ref{C211eq}), (\ref{C212eq})
two--point WF $G_{_0 (\0\1)}$ and $D_{_0 (\0\1)}$ calculated in
Hartree--Fock approximation are used.

As accurate within first order of vanishing terms $\di\la v^2=
\frac{N}{8}m_1^2\;$, corrections to WF $G_{_1 (\0\1)}$ and $D_{_1
(\0\1)}$ are of the first order of vanishing and should be taken
into account along with $C_{_{03} (\1\2\3)}$ and $C_{_{21}
(\1\2|\3)}$. Thus, at second step equations for $G_{_1 (\0\1)}$
and $D_{_1 (\0\1)}$, which derived from (\ref{Gequation}) and
(\ref{Dequation}) by substitution $\la v^2$, expressed through
$m_1^2$, substitution of three-point WF and neglecting of the
second order of vanishing sources (in particular four--point WF,
multiplied by $\la$).
\begin{equation}
\partial_\mu\partial^\mu_{x}G_{_1 (\0\1)}+m_1^2\; G_{_1
(\0\1)}\approx-\frac{m_1^2}{2}\left[3C_{_{03}(\0\0\1)}+
(N-1)C_{_{21}(\0\0|\1)}\right], \label{G1}
\end{equation}
\begin{equation}
\di \partial_\mu\partial^\mu_{x}D_{_1 (\0\1)}+m_2^2\; D_{_1
(\0\1)}\approx -m_1^2\; C_{_{21}(\0\1|\0)}. \label{D1}
\end{equation}
Taking into account tree-point WF and corrections to two--point WF
in equation for order parameter and equations for effective masses
leads to new temperature dependence of these values in compare
with Hartree--Fock approximation. New dependencies are found at
third step.

Preliminary conclusion of corrections to $m_1$, $m_2$ and $v$
magnitudes one could made without numerical analysis. Assuming the
corrections small, introduce effective masses and order parameter
as follows:
\begin{equation}
m_1^2=m_{1 0}^2+\delta m_1^2,\; m_2^2=m_{2 0}^2+\delta m_2^2,\;
v^2=v_0^2+\delta v^2, \label{pres}
\end{equation}
Here $m_{1 0}$, $m_{2 0}$ and $v_0$ corresponds to self consistent
field approximation. Substitution of (\ref{pres}) in
(\ref{O(N)condensat_2}) and  (\ref{O(N) m_1}), (\ref{O(N) m_2})
and separation by orders of vanishing gives the following
expressions for corrections:
\begin{equation}
\begin{array}{c}
\di\delta m_2^2=-\frac{4\la (N+2)}{\delta\cdot N}\biggl[
2(N-1)D_{_1(\0\0)}-\left(\frac{4\la(N+2)}{N}f_1(m_{2
0},T)-3\right)\cdot
\\[4mm]
\cdot\left(2
G_{_1(\0\0)}+C_{_{03}(\0\0\0)}+(N-1)C_{_{21}(\0\0|\0)} \right)
\biggr], \label{deltam1}
\end{array}
\end{equation}
\begin{equation}
\begin{array}{c}
\di\delta m_2^2=-\frac{4\la (N+2)}{\delta\cdot N}\biggl[-2
D_{_1(\0\0)}\left(\frac{4\la(N+2)}{N}f_1(m_{1 0},T)+1\right)+2
G_{_1(\0\0)}+
\\[4mm]\di
+C_{_{03}(\0\0\0)}+(N-1)C_{_{21}(\0\0|\0)} \biggr],
\label{deltam2}
\end{array}
\end{equation}
\begin{equation}
\delta v^2=\frac{\delta m_1^2\cdot
N}{8\la}+\frac12\left(C_{_{03}(\0\0\0)}+(N-1)C_{_{21}(\0\0|\0)}\right).
\label{deltav}
\end{equation}
Designations $f_1(m,T)$, which up to a coefficient is a squared
mass derivative of two--point WF considered in Hartree--Fock
approximation,
\[f_1(m_{1 0},T)=2\frac{\partial G_{_0 (\0\0)}}{\partial m_{1
0}^2},\; f_1(m_{2 0},T)=2\frac{\partial D_{_0 (\0\0)}}{\partial
m_{2 0}^2}
\]
and $\delta$ were introduced in chapter \ref{PSP}. As long as at
the vicinity of $T_{c_2}$ $\delta\sim\sqrt{T_{c_2}-T}$,
corrections, calculated by (\ref{deltam1}), (\ref{deltam2}) and
(\ref{deltav}), are of high magnitude in this region.

At fourth step corrections to thermodynamical observables are
calculated. Neglecting of four--point Wightman functions, which
contributions are of second order of vanishing, and substitution
of two-- and three--point WF, expressed through  effective masses
and condensate value, in ``potential'' terms of momentum--energy
tensor gives:
\begin{equation}
\begin{array}{c}
\di \langle T_\mu^\nu\rangle=\lim_{x\rightarrow
x_1}\left(\partial_{\mu\hspace*{0.5mm}
(x)}\partial^\nu_{(x_{1})}G_{(x
x_1)}+(N-1)\partial_{\mu\hspace*{0.5mm}
(x)}\partial^\nu_{(x_{1})}D_{(x x_1)}\right)-
\\[4mm]\di
-\delta_\mu^\nu\biggl(\frac{N}{8\la (N+2)}\left[\di
\frac{(N+1)m_1^4+3(N-1)m_2^4}{4}-\frac{N-1}{2}\hspace*{1mm} m_1^2
\hspace*{0.5mm} m_2^2+\right.
\\[4mm]\di
+\mu^2(m_1^2+(N-1)m_2^2)+\frac{N\mu^4}{2}\biggr]-m_1^2
v^2+\frac{6\la v^4}{N}\biggr). \label{newTEI}
\end{array}
\end{equation}
Since in high--symmetry phase three--point WF are identically
equal to zero, corrections are of the second order of vanishing,
and should not be considered in approximation which takes into
account only effects of zero and first order of vanishing.

\paragraph{Calculation of three--point WF.}
In Hartree--Fock approximation two--point WF $G_{(\0\1)}$ and
$D_{(\0\1)}$, and commutators look like:
\[ \di
G_{_0(\0\1)}=\sum_\vecp \frac{n_\vecp+\frac12}{2 \ve_\vecp
V}\left(e^{ip(\0-\1)}+e^{ip(\1-\0)}\right),\;D_{_0(\0\1)}=\sum_\veck
\frac{n_\veck+\frac12}{2 \ve_\veck
V}\left(e^{ik(\0-\1)}+e^{ik(\1-\0)}\right), \] \[
\Delta\vp_{_0(\0\1)}=\sum_\vecp \frac1{2 \ve_\vecp
V}\left(e^{ip(\0-\1)}-e^{ip(\1-\0)}\right),\;\Delta\chi_{_0(\0\1)}=\sum_\veck
\frac1{2 \ve_\veck V}\left(e^{ik(\0-\1)}-e^{ik(\1-\0)}\right), \]
so that sources in equations (\ref{C03eq}), (\ref{C211eq}),
(\ref{C212eq}) and (\ref{G1}), (\ref{D1}) are sum of exponents.
Since exponent is an eigen-function of differentiation operator,
action of D'Alamber operator at each terms reduce to multiplying
by correspondent function of masses and impulses $f(q,m)$. Hence,
solutions of the equations are functions which differ from sources
by factors $f(q,m)^{-1}$ of exponents.

Use of this algorithm and consideration of symmetrical properties
of Wightman functions $C_{_{03}(\0\1\2)}$ gives:
\begin{equation}
\begin{array}{c}
\di \la C_{_{03}(\0\1\2)}=\lim_{n\rightarrow 3
}\frac{-12\la_n^2}{N (2\pi)^{2 n}}\int\frac{d^n p\hspace*{1mm} d^n
p'}{\ve_p \ve_{p'}}\biggl[\left(\frac{1}{\exp{\frac{\ve_p}{T}} -
1} +\frac12\right) \biggl(\frac{1}{\exp{\frac{\ve_{p'}}{T}} - 1}
+\frac12\biggr)\times
\\[5mm]
\di {\rm Re}\left(\frac{ e^{i{\rm p}(x-x_1)}e^{i{\rm
p}'(x-x_2)}+e^{i{\rm p}(x_1-x)}e^{i{\rm p'}(x_1-x_2)}+ e^{i{\rm
p}(x_2-x)}e^{i{\rm p'}(x_2-x_1)} }{m_1^2-({\rm p}+{\rm
p'})^2}\;+\right.
\\[5mm]
\di+ \left.\frac{ e^{i{\rm p}(x-x_1)}e^{i{\rm p'}(x_2-x)}+e^{i{\rm
p}(x_1-x)}e^{i{\rm p'}(x_2-x_1)}+ e^{i{\rm p}(x_2-x)}e^{i{\rm
p'}(x_1-x_2)} }{m_1^2-({\rm p}-{\rm p'})^2} \right)+
\end{array}
\label{C03pribl}
\end{equation}
\[
\begin{array}{c}
\di +\frac1{12}\cdot{\rm Re}\left(\frac{ e^{i{\rm
p}(x-x_1)}e^{i{\rm p'}(x-x_2)}+e^{i{\rm p}(x_1-x)}e^{i{\rm
p'}(x_1-x_2)}+ e^{i{\rm p}(x_2-x)}e^{i{\rm p'}(x_2-x_1)}
}{m_1^2-({\rm p}+{\rm p'})^2}\;-\right.
\\[5mm]
\di- \left.\left.\frac{ e^{i{\rm p}(x-x_1)}e^{i{\rm
p'}(x_2-x)}+e^{i{\rm p}(x_1-x)}e^{i{\rm p'}(x_2-x_1)}+ e^{i{\rm
p}(x_2-x)}e^{i{\rm p'}(x_1-x_2)} }{m_1^2-({\rm p}-{\rm p'})^2}
\right)\right].
\end{array}
\]
Here $\di {\rm
p}=(p^{_0},\vecp)=(\sqrt{\vecp\hspace*{1mm}^2+m_1^2},\vecp)$, $\di
{\rm p'}=(p'^{_0},\vecp')=(\sqrt{\vecp\hspace*{0.5mm}
'\hspace*{0.5mm}^2+m_1^2},\vecp\hspace*{0.5mm}')$. To obtain
(\ref{C03pribl}) â $G_{_0(\0\1)}$, $D_{_0(\0\1)}$, $\Delta
\vp_{_0(\0\1)}$ and $\Delta \chi_{_0(\0\1)}$ one should turn from
summation to integration over non--integer dimension space.

In solutions of (\ref{C211eq}) and  (\ref{C212eq}) terms $\di
\frac1{m_2^2-({\rm p}-{\rm k})^2}$ and $\di\frac1{m_1^2-({\rm
k}+{\rm k'})^2}$ respectively arise. Denominators of these terms
turn to zero for some impulses $\vecp$, $\veck$ if condition
$m_1\geq 2 m_2 $ is satisfied. Physically satisfaction of this
condition corresponds to possibility of $\vp$ to pair $\chi_a$
decay. In tree approximation decay width is:
\[ \di
\Gamma=\frac{N-1}{N^2}\cdot\frac{2\la^2 v^2}{\pi
m_1}\sqrt{1-\left(\frac{2 m_2}{m_1}\right)^2}. \] Let us now
introduce helper function $C^{\;au}_{_{ 21} (\0\1|\2)}$, which
satisfy to equations obtained from (\ref{C211eq}) and
(\ref{C212eq}) respectively by substitutions
\[
\partial_\mu\partial^\mu_{x_2}+m_1^2\Rightarrow
\partial_\mu\partial^\mu_{x_2}+m_1^2-i m_1\Gamma,\hspace*{5mm}
\partial_\mu\partial^\mu_{x}+m_2^2\Rightarrow
\partial_\mu\partial^\mu_{x}+m_2^2-i m_2\Gamma.
\]
Since WF are real, by definition $\di C_{_{21}
(\0\1|\2)}=\frac12\left[C^{\;au}_{_{21}
   (\0\1|\2)}+C^{\;au\;*}_{_{21}
   (\0\1|\2)}\right]$.

\noindent After consideration of $C_{_{21}(\0\1|\2)}$ symmetries
with regard to arguments permutation:
\begin{equation}
\begin{array}{c} \di \la C_{_{21}(\0\1|\2)}=\lim_{n\rightarrow 3
}\frac{-4\la_n^2}{N (2\pi)^{2n}}\int\frac{d^n p\hspace*{1mm} d^n
k}{\ve_\vecp \ve_\veck}\left(\frac{1}{\exp{\frac{\ve_\vecp}{T}} -
1} +\frac12\right) \left(\frac{1}{\exp{\frac{\ve_\veck}{T}} - 1}
+\frac12\right)\times \\[5mm]\di{\rm Re} \left(\frac{{\rm
Re}\left(e^{i{\rm {\rm p}}(x-x_1)}e^{i{\rm k}(x-x_2)}+e^{i{\rm
p}(x_1-x)}e^{i{\rm k}(x_1-x_2)}\right)} {m_2^2-({\rm p}+{\rm
k})^2-im_2\Gamma}\;+\hspace*{50mm}\right.
\\[2mm]\di\hspace*{50mm}+\left. \frac{{\rm Re}\left(e^{i{\rm
p}(x-x_1)}e^{i{\rm k}(x_2-x)} +e^{i{\rm p}(x_1-x)}e^{i{\rm
k}(x_2-x_1)}\right)}{m_2^2-({\rm p}-{\rm k})^2-i m_2\Gamma }
\right)- \\[4mm]\di
\end{array}
\end{equation}
\[
\begin{array}{c}\di
-\lim_{n\rightarrow 3 }\frac{-4\la_n^2}{N(2\pi)^6}\int\frac{d^n
k\hspace*{1mm}
 d^n k'}{\ve_k
\ve_{k'}}\biggl[\left(\frac{1}{\exp{\frac{\ve_k}{T}} - 1}
+\frac12\right) \left(\frac{1}{\exp{\frac{\ve_{k'}}{T}} - 1}
+\frac12\right)\times\\[5mm]\hspace*{10mm} \di{\rm Re} \left(\frac{{\rm Re
}\left(e^{i{\rm k}(x_2-x)}e^{i{\rm k'}(x_2-x_1)}\right)
}{m_1^2-({\rm k}+{\rm k'})^2-im_1\Gamma}+\frac{{\rm
Re}\left(e^{i{\rm k}(x-x_2)}e^{i{\rm k'}(x_2-x_1)}\right)}
{m_1^2-({\rm k}-{\rm
k'})^2-im_1\Gamma} \right)+ \\[5mm]\di+\left. \frac1{12}\cdot \di{\rm Re}
\left(\frac{{\rm Re }\left(e^{i{\rm k}(x_2-x)}e^{i{\rm
k}'(x_2-x_1)}\right) }{m_1^2-({\rm k}+{\rm
k}')^2-im_1\Gamma}-\frac{{\rm Re}\left(e^{i{\rm k}(x-x_2)}e^{i{\rm
k}'(x_2-x_1)}\right)} {m_1^2-({\rm k}-{\rm k}')^2-im_1\Gamma}
\right)\right]. \end{array} \label{C21pribl} \] Here $\di {\rm
k}=(k^{_0},\veck)=(\sqrt{\veck\hspace*{1mm}^2+m_1^2},\veck)$, $\di
{\rm k'}=(k'^{_0},\veck')=(\sqrt{\veck\hspace*{0.5mm}
'\hspace*{0.5mm}^2+m_1^2},\veck\hspace*{0.5mm}')$. Solutions,
written above, are symmetrical of permutations of arguments. Taken
in coincide points, they depend on inner system parameters, but
not 4--coordinates.

Wightman functions $C_{_{03} (\0\0\0)}$ and $C_{_{21} (\0\0|\0)}$
automatically separate into items of three types. Items of the
first type contain product of Bose--Einstein distributions under
the integral and are finite at the limit $n\rightarrow 3$. Items
of the second type (``mixed integrals'') which contain under
integral Bose--Einstein distribution for one integration variable
and items of the third type (``vacuum integrals'') which don't
contain Bose--Einstein distribution under integral are divergent
at the limit $n\rightarrow 3$.

Asymptotic expansion of integrands is used for regularization of
divergent terms of three--point WF in the framework of dimensional
regularization. Three--dimensional impulse $\vecp$ ($\veck$) to
effective mass $m_1$ ($m_2$) quotient serves as expansion
parameter, so that region of integration is divided into four
regions where values of these parameters are greater (less) than
one. In each region an expansion valid in the region is used.

The integrals are well defined in space of dimension $1<n<2$.
Since ``vacuum integrals'' are divergent by both integration
variables, analytical continuation into $n=3$ dimension space lead
not only to $\ve^{-1}$ pole (which is responsible for $\di
\ln\left(\frac{m^2}{\Lambda^2}\right)$ terms in renormalized
integrals), but also $\ve^{-2}$ pole (which is responsible for
$\di\ln^2\left(\frac{m^2}{\Lambda^2}\right)$ terms).

As ``mixed integrals'' are divergent only by one integration
variable, at the limit $n\rightarrow 3$ only $\ve^{-1}$ pole
arise. Coefficient at $\di \ln\left(\frac{m^2}{\Lambda^2}\right)$,
formed by Bose--Einstein distribution, is a function of particle
mass and temperature.

After poles subtraction one obtain the following expressions for
three--point Wightman functions:
\[ \begin{array}{c} \di C_{03
(\0\0\0)_{ren}}=\frac{-18\la}{(2\pi)^2
N}\left[I_1^0(m_1,m_1,m_1,T)+J_1(m_1,T)\left({\bf C_1}-2
\ln\left(\frac{m_1^2}{\Lambda^2}\right)\right)+\right. \\[3mm]\di
\left.+K_1^0(m_1,m_1,m_1,T)+{\bf C_2}\hspace*{0.5mm}
m_1^2\hspace*{0.5mm}
 \ln^2\left(\frac{m_1^2}{\Lambda^2}\right)+{\bf C_3}\hspace*{0.5mm}m_1^2\hspace*{0.5mm}
\ln\left(\frac{m_1^2}{\Lambda^2}\right)+{\bf
C\hspace*{0.5mm}_4}\hspace*{0.5mm}m_1^2\right], \end{array} \] \[
\begin{array}{c}\di C_{21 (\0\0|\0)_{ren}}=\frac{-4\la}{(2\pi)^2
N}\left[I_1^0(m_1,m_2,m_2,T)+\frac12
\left(J_1(m_1,T)+\frac{m_1^2}{m_2^2} J_1(m_2,T)\right)\biggl({\bf
C_1}-\right. \\[4mm]\di
\left.\left. -2\ln\left(\frac{m_1\hspace*{0.5mm} m_2
}{\Lambda^2}\right)\right)+ \frac12 K_1^0(m_1,m_2,m_2,T)+\frac12
K_1^0(m_2,m_1,m_2,T)+ {\bf C_2}\hspace*{0.5mm}
m_1^2\hspace*{0.5mm}
 \ln^2\left(\frac{m_1\hspace*{0.5mm} m_2}{\Lambda^2}\right)+\right.
\\[4mm]\di +{\bf C_3}\hspace*{0.5mm}m_1^2\hspace*{0.5mm}
\ln\left(\frac{m_1\hspace*{0.5mm} m_2}{\Lambda^2}\right)+{\bf
C\hspace*{0.5mm}_5}\hspace*{0.5mm}m_1^2+
L_1^0(m_1,m_2,m_2,1)\biggr] + \frac{-2\la}{(2\pi)^2 N
}\left[I_1^0(m_2,m_2,m_1,T)+\right. \\[3mm]\di
\left.+K_1^0(m_2,m_2,m_1,T) +\left(2-\frac{m_1^2}{m_2^2}\right)
J_1(m_2,T)\left({\bf
C_1}-2\ln\left(\frac{m_2^2}{\Lambda^2}\right)\right)-{\bf
C\hspace*{0.5mm}_5}\hspace*{0.5mm}m_1^2+{\bf
C\hspace*{0.5mm}_6}\hspace*{0.5mm}m_2^2+\right. \\[4mm]
\di
\end{array}
\]
\[
\begin{array}{c}
\left.+ {\bf C_2}\left(2
m_2^2-m_1^2\right)\ln^2\left(\frac{m_2^2}{\Lambda^2}\right) +{\bf
C_3}\left(2
m_2^2-m_1^2\right)\ln\left(\frac{m_2^2}{\Lambda^2}\right)+L_1^0(m_2,m_2,m_1,4/3)
\right. \biggr]. \end{array} \]

\noindent Coefficients $C_i$ were calculated approximately: $\di
{\bf C_1}\approx-1.386, {\bf C_2}\approx 0.044, {\bf C_3}\approx
0.013, {\bf C_4}\approx -0.031, {\bf C_5}\approx -0.012,{\bf
C_5}\approx -0.031, {\bf C_6}\approx -0.089$. Introduced above
functions $I_n(x,y,\alpha,T)$, $K_n(x,y,\alpha,T)$ and
$L_n(x,y,\alpha,\beta,T)$ are defined as follows:
\begin{equation} \begin{array}{c} \di
I_n^m(x,y,\alpha,T)=\frac2{(2\pi)^4}\int\int
\frac{d\hspace*{0.5mm}^3p\; d\hspace*{0.5mm}^3k}{\ve_\vecp\;
\ve_\veck }\cdot\di\frac{1}{\exp{\frac{\ve_\vecp}{T}}-1}\cdot
\frac{1}{\exp{\frac{\ve_\veck}{T}}-1}\;\cdot \\[3mm]\di\cdot{\rm Re
}\left[\frac{(\vecp+\veck)^{2m}}{\left(\alpha^2-({\rm p}+{\rm
k})^2-i\alpha\Gamma\right)^n}\hspace*{1mm}
+\frac{(\vecp-\veck)^{2m}} {\left(\alpha^2-({\rm p}-{\rm
k})^2-i\alpha\Gamma\right)^n} \right], \end{array} \label{I1}
\end{equation} \begin{equation} \begin{array}{c} \di
K_n^m(x,y,\alpha,T)=\frac2{(2\pi)^4}\int_{|\vecp|=0}^{|\vecp|=\infty}
\int_{|\veck|=0}^{|\veck|=y} \frac{d\hspace*{0.5mm}^3p\;
d\hspace*{0.5mm}^3k}{\ve_\vecp\; \ve_\veck
}\cdot\di\frac{1}{\exp{\frac{\ve_\vecp}{T}}-1}\;\cdot \\[3mm]\di
\cdot{\rm Re }\left[\frac{(\vecp+\veck)^{2m}}{\left(\alpha^2-({\rm
p}+{\rm k})^2-i\alpha\Gamma\right)^n}\hspace*{1mm}
+\frac{(\vecp-\veck)^{2m}} {\left(\alpha^2-({\rm p}-{\rm
k})^2-i\alpha\Gamma\right)^n} \right], \end{array} \label{K1}
\end{equation} \begin{equation} \begin{array}{c} \di
L_n^m(x,y,\alpha,\beta)=\frac1{2
(2\pi)^4}\int_{|\vecp|=0}^{|\vecp|=x} \int_{|\veck|=0}^{|\veck|=y}
\frac{d\hspace*{0.5mm}^3p\;
d\hspace*{0.5mm}^3k}{\ve_\vecp\; \ve_\veck }\;\cdot \\[3mm]\cdot
\di {\rm Re
}\left[\frac{\beta(\vecp+\veck)^{2m}}{\left(\alpha^2-({\rm p}+{\rm
k})^2-i\alpha\Gamma\right)^n}\hspace*{1mm} +\frac{(2-\beta)
(\vecp-\veck)^{2m}}{\left(\alpha^2-({\rm p}-{\rm
k})^2-i\alpha\Gamma\right)^n} \right]. \end{array} \label{L1}
\end{equation} In (\ref{I1}),(\ref{K1}) and (\ref{L1})
4--impulses ${\rm p}$ and ${\rm k}$ depend on arguments $x$ and
$y$ as:
$\di {\rm
p}=(p^{_0},\vecp)=(\ve_\vecp,\vecp)=(\sqrt{\vecp\hspace*{1mm}^2+x^2},\vecp)$,
${\rm
k}=(k^{_0},\veck)=(\ve_\veck,\veck)=(\sqrt{\veck\hspace*{1mm}^2+x^2},\veck)$.

Proportional to introduced in chapter \ref{PSP} temperature
integral $J_1(m,T)$ terms are renormalized ``mixed integrals''.
Renormalized ``vacuum integrals'' are not explicitly temperature
dependent.

Wightman function $C_{21 (\0\0|\0)_{ren}}$, which contain terms
$\di m_1^2\hspace*{0.5mm}\ln\left(\frac{m_2^2}{\Lambda^2}\right)$
and $\di m_1^2\hspace*{0.5mm}\ln\left(\frac{m_1\hspace*{0.5mm}
m_2}{\Lambda^2}\right)$ is logarithmically divergent at the limit
$T\rightarrow 0$ . This divergence, caused by long-range action,
is described by diagrams of massless $\chi_a$ quanta exchange.
Since at the most interesting region of phase transition $m_2$ is
not zero, this problem was not investigated in details.

\paragraph{Calculation of corrections to two--point Wightman functions.}
Taking into account (\ref{C03equation}), (\ref{C211equation}) and
(\ref{C212equation}) it is convenient to rewrite equations
(\ref{G1}) and (\ref{D1}) as follows:\begin{equation}
\begin{array}{c} \di
\left[\partial_\mu\partial^\mu_{\1}+m_1^2\right]\left[\partial_\mu\partial^\mu_{\0}+m_1^2\right]G_{_1(\0\1)}=
\frac{36\la\;m_1^2}{N}\left[G_{_0(\0\1)}G_{_0(\0\1)}+\frac1{12}\Delta
\vp_{_0(\0\1)}\Delta \vp_{_0(\0\1)}\right]+ \\[4mm]\di
+\frac{4\la m_1^2
(N-1)}{N}\left[D_{_0(\0\1)}D_{_0(\0\1)}+\frac1{12}\Delta
\chi_{_0(\0\1)}\Delta \chi_{_0(\0\1)}\right], \label{G1eqnew}
\end{array} \end{equation} \begin{equation} \di
\left[\partial_\mu\partial^\mu_{\1}+m_2^2\right]\left[\partial_\mu\partial^\mu_{\0}+m_2^2\right]D_{_1(\0\1)}=
\frac{8\la\;m_1^2}{N}D_{_0(\0\1)}G_{_0(\0\1)}. \label{D1eqnew}
\end{equation} Since, like in the case of three--point WF, divergent
for $m_1>2m_2$ and some impulses  terms $\di \frac1{(m_2^2-({\rm
p}-{\rm k})^2)^2}$ and $\di\frac1{(m_1^2-({\rm k}+{\rm k'})^2)^2}$
arise in solutions, equations for two--point WF should be
redefined. Use of the same way of redefinition as was used for
$C_{_{21}(\0\1|\2)}$ leads to:
\begin{equation} \begin{array}{c}
\di\la G_{_1 (\0\1)}=\lim_{n\rightarrow 3}\frac{2\la_n^2
m_1^2}{N(2\pi)^{2n}}\biggl[9\int\frac{d^n p\hspace*{1mm} d^n
p'}{\ve_p \ve_{p'}}\biggl[\left(\frac{1}{\exp{\frac{\ve_p}{T}} -
1} +\frac12\right) \biggl(\frac{1}{\exp{\frac{\ve_{p'}}{T}} - 1}
+\frac12\biggr)\times\\[5mm]\di
\end{array}
\end{equation}
\[
\begin{array}{c}\di
{\rm Re}\left(\frac{{\rm Re}(e^{i{\rm p}(\1-\0)}e^{i{\rm
p}'(\1-\0)}) }{(m_1^2-({\rm p}+{\rm
p}')^2-im_1\Gamma)^2}+\frac{{\rm Re}(e^{i{\rm p}(\1-\0)}e^{i{\rm
p}'(\0-\1)})}{(m_1^2-({\rm p}-{\rm
p}')^2-im_1\Gamma)^2} \right)+ \\[4mm]\di \frac1{12}\cdot{\rm
Re}\left(\frac{{\rm Re}(e^{i{\rm p}(\1-\0)}e^{i{\rm p}'(\1-\0)})
}{(m_1^2-({\rm p}+{\rm p}')^2-im_1\Gamma)^2}\;-\frac{{\rm
Re}(e^{i{\rm p}(\1-\0)}e^{i{\rm p}'(\0-\1)}) }{(m_1^2-({\rm
p}-{\rm p}')^2-im_1\Gamma)^2}\right)\biggr]+ \\[4mm]\di
+(N-1)\int\frac{d^n k\hspace*{1mm} d^n k'}{\ve_\veck
\ve_{\veck\;'}}\biggl[\left(\frac{1}{\exp{\frac{\ve_\veck}{T}} -
1} +\frac12\right) \biggl(\frac{1}{\exp{\frac{\ve_{\veck'}}{T}} -
1} +\frac12\biggr)\times \\[5mm]\di {\rm Re}\left(\frac{{\rm
Re}(e^{i{\rm k}(\1-\0)}e^{i{\rm k}'(\1-\0)})}{(m_1^2-({\rm k}+{\rm
k}')^2-im_1\Gamma)^2}+\frac{{\rm Re}(e^{i{\rm k}(\1-\0)}e^{i{\rm
k}'(\0-\1)}) }{(m_1^2-({\rm k}-{\rm
k}')^2-im_1\Gamma)^2} \right)+ \\[4mm]\di +\frac1{12}\cdot{\rm
Re}\left(\frac{{\rm Re}(e^{i{\rm k}(\1-\0)}e^{i{\rm
k}'(\1-\0)})}{(m_1^2-({\rm k}+{\rm
k}')^2-im_1\Gamma)^2}\;-\frac{{\rm Re}(e^{i{\rm k}(\1-\0)}e^{i{\rm
k}'(\0-\1)})}{(m_1^2-({\rm k}-{\rm
k}')^2-im_1\Gamma)^2}\right)\biggr] \end{array}\label{G1xx}
\] \begin{equation} \begin{array}{c} \di\la D_{_1
(\0\1)}=\lim_{n\rightarrow 3}\frac{4\la_n^2 m_1^2}{N
(2\pi)^{2n}}\int\frac{d^n p\hspace*{1mm} d^n k}{\ve_\vecp
\ve_\veck}\left(\frac{1}{\exp{\frac{\ve_\vecp}{T}} - 1}
+\frac12\right) \left(\frac{1}{\exp{\frac{\ve_\veck}{T}} - 1}
+\frac12\right)\times \\[5mm]\di {\rm Re}\left(\frac{{\rm
Re}(e^{i{\rm p}(\1-\0)}e^{i{\rm k}(\1-\0)})} {(m_2^2-({\rm p}+{\rm
k})^2-im_2\Gamma)^2}+\frac{{\rm Re}(e^{i{\rm p}(\1-\0)}e^{i{\rm
k}(\0-\1)})} {(m_2^2-({\rm p}-{\rm k})^2-im_2\Gamma)^2} \right).
\end{array} \label{D1xx}
\end{equation}
After renormalization of corrections to two--point WF, considered
in coincide points, in the framework of dimensional regularization
one obtain:
\[ \begin{array}{c} \di
G_{_1(\0\0)_{ren}}=\frac{9\la}{N(2\pi)^2}\left[ m_1^2\;
I_2^0(m_1,m_1,m_1,T)+J_1(m_1,T)\left(C_1-2\ln\left(\frac{m_1^2}{\Lambda^2}\right)\right)\right.+
\\[4mm]\di
\left.+m_1^2\;K_2^0(m_1,m_1,m_1,T)+C_2\;m_1^2\ln^2\left(\frac{m_1^2}{\Lambda^2}\right)+C_7\;m_1^2
\ln\left(\frac{m_1^2}{\Lambda^2}\right)+C_8\;m_1^2\right]+
\\[4mm]\di +\frac{(N-1)\la}{N(2\pi)^2}\left[ m_1^2\;
I_2^0(m_2,m_2,m_1,T)+\frac{m_1^2}{m_2^2}J_1(m_2,T)
\left(C_1-2\ln\left(\frac{m_2^2}{\Lambda^2}\right)
\right)+C_{9}\;m_1^2+C_{10}\;m_2^2+\right.
\\[4mm]\di\left.+m_1^2\;K_2^0(m_2,m_2,m_1,T)+C_2\;m_1^2\ln^2
\left(\frac{m_2^2}{\Lambda^2}\right) +C_7\;m_1^2
\ln\left(\frac{m_2^2}{\Lambda^2}\right)+m_1^2\;L_2^0(m_2,m_2,m_1,4/3)
\right], \end{array} \] \[ \begin{array}{c} \di
D_{_1(\0\0)_{ren}}=\frac{2\la}{N(2\pi)^2}\left[
m_1^2\;I_2^0(m_1,m_2,m_2,T)+\left(J_1(m_1,T)+\frac{m_1^2}{m_2^2}J_2(m_2,T)\right)
\left(\frac{C_1}{2}\;-\right.\right. \\[4mm]\di
\left.-\ln\left(\frac{m_1\;m_2}{\Lambda^2}\right)\right)+\frac{m_1^2}2
K_2^0(m_1,m_2,m_2,T)+\frac{m_1^2}2
K_2^0(m_2,m_1,m_2,T)+C_2\;m_1^2\ln^2\left(\frac{m_1\;m_2}{\Lambda^2}\right)+
\\[4mm]\di +C_7\;m_1^2
\ln\left(\frac{m_1\;m_2}{\Lambda^2}\right)+C_{11}\;m_1^2+L_2^0(m_1,m_2,m_2,1)
\biggl]. \end{array} \] Here $\di {\bf C_7}\approx-0.568, {\bf
C_8}\approx-0.929, {\bf C_9}\approx -0.111, {\bf C_{10}}\approx
-0.818, {\bf C_{11}}\approx -1.042$.

In order to calculate (specific) equilibrium free energy, which is
expressed via momentum--energy tensor components by relation
(\ref{omega-pot}), considering correction to Hartree--Fock
approximation of the first order of vanishing, one should derive
and renormalize derivatives of corrections to two--point WF.

Differentiation of (\ref{G1xx}) and (\ref{D1xx}) gives
respectively:
\[
\begin{array}{c} \di \lim_{n\rightarrow
3}\frac{V_n}{V}\frac1{n}\cdot\partial_{i\hspace*{0.5mm}
(x)}\partial^i_{(x_{1})}G_{_1(x x_1)}= \\[4mm]\di
=-\lim_{n\rightarrow 3}\frac1{n}\cdot\frac{2\la_n^2
m_1^2}{N(2\pi)^{2n}}\biggl[9\int\frac{d^n p\hspace*{1mm} d^n
p'}{\ve_p \ve_{p'}}\biggl[\left(\frac{1}{\exp{\frac{\ve_p}{T}} -
1} +\frac12\right) \biggl(\frac{1}{\exp{\frac{\ve_{p'}}{T}} - 1}
+\frac12\biggr)\times \\[5mm]\di
\end{array}
\]
\[
\begin{array}{c}\di
{\rm Re}\left(\frac{(\vecp+\vecp')^2\cdot{\rm Re}(e^{i{\rm
p}(\1-\0)}e^{i{\rm p}'(\1-\0)}) }{(m_1^2-({\rm p}+{\rm
p}')^2-im_1\Gamma)^2}+\frac{(\vecp-\vecp')^2\cdot{\rm Re}(e^{i{\rm
p}(\1-\0)}e^{i{\rm p}'(\0-\1)})}{(m_1^2-({\rm
p}-{\rm p}')^2-im_1\Gamma)^2} \right)+ \\[4mm]\di
\frac1{12}\cdot{\rm Re}\left(\frac{(\vecp+\vecp')^2\cdot{\rm
Re}(e^{i{\rm p}(\1-\0)}e^{i{\rm p}'(\1-\0)}) }{(m_1^2-({\rm
p}+{\rm p}')^2-im_1\Gamma)^2}\;-\frac{(\vecp-\vecp')^2\cdot{\rm
Re}(e^{i{\rm p}(\1-\0)}e^{i{\rm p}'(\0-\1)}) }{(m_1^2-({\rm
p}-{\rm p}')^2-im_1\Gamma)^2}\right)\biggr]+ \\[4mm]\di
+(N-1)\int\frac{d^n k\hspace*{1mm} d^n k'}{\ve_\veck
\ve_{\veck\;'}}\biggl[\left(\frac{1}{\exp{\frac{\ve_\veck}{T}} -
1} +\frac12\right) \biggl(\frac{1}{\exp{\frac{\ve_{\veck'}}{T}} -
1} +\frac12\biggr)\times \\[5mm]\di {\rm
Re}\left(\frac{(\veck+\veck')^2\cdot{\rm Re}(e^{i{\rm
k}(\1-\0)}e^{i{\rm k}'(\1-\0)})}{(m_1^2-({\rm k}+{\rm
k}')^2-im_1\Gamma)^2}+\frac{(\veck-\veck')^2\cdot{\rm Re}
(e^{i{\rm k}(\1-\0)}e^{i{\rm k}'(\0-\1)}) }{(m_1^2-({\rm k}-{\rm
k}')^2-im_1\Gamma)^2} \right)+ \\[4mm]\di +\frac1{12}\cdot{\rm
Re}\left(\frac{(\veck+\veck')^2\cdot{\rm Re}(e^{i{\rm
k}(\1-\0)}e^{i{\rm k}'(\1-\0)})}{(m_1^2-({\rm k}+{\rm
k}')^2-im_1\Gamma)^2}\;-\frac{(\veck-\veck')^2\cdot{\rm
Re}(e^{i{\rm k}(\1-\0)}e^{i{\rm k}'(\0-\1)})}{(m_1^2-({\rm k}-{\rm
k}')^2-im_1\Gamma)^2}\right)\biggr], \end{array} \] \[
\begin{array}{c} \di \lim_{n\rightarrow
3}\frac{V_n}{V}\frac1{n}\cdot\partial_{i\hspace*{0.5mm}
(x)}\partial^i_{(x_{1})}D_{_1(x x_1)}= \\[4mm]\di
=-\lim_{n\rightarrow 3}\frac1{n}\cdot\frac{4\la_n^2 m_1^2}{N
(2\pi)^{2n}}\int\frac{d^n p\hspace*{1mm} d^n k}{\ve_\vecp
\ve_\veck}\left(\frac{1}{\exp{\frac{\ve_\vecp}{T}} - 1}
+\frac12\right) \left(\frac{1}{\exp{\frac{\ve_\veck}{T}} - 1}
+\frac12\right)\times \\[5mm]\di {\rm
Re}\left(\frac{(\vecp+\veck)^2\cdot{\rm Re}(e^{i{\rm
p}(\1-\0)}e^{i{\rm k}(\1-\0)})} {(m_2^2-({\rm p}+{\rm
k})^2-im_2\Gamma)^2}+\frac{(\vecp-\veck)^2\cdot{\rm Re}(e^{i{\rm
p}(\1-\0)}e^{i{\rm k}(\0-\1)})} {(m_2^2-({\rm p}-{\rm
k})^2-im_2\Gamma)^2} \right). \end{array} \] After renormalization
one obtain:
\[
\begin{array}{c} \di \lim_{\1\rightarrow\0}
G^{''}_{_1(\0\1)_{ren}}\equiv\left(\lim_{\1\rightarrow\0}\frac1{n}
\frac{V_n}{V}\cdot\partial_{i\hspace*{0.5mm}
(x)}\partial^i_{(x_{1})}G_{_1(x
x_1)}\right)_{ren}=-\frac13\cdot\frac{\la
m_1^2}{N(2\pi)^2}\biggl[ 9\biggl(I^1_2(m_1,m_1,m_1,T)+ \\[4mm]\di
+K^1_2(m_1,m_1,m_1,T)-
 \frac12 J_1(m_1,T)+\frac1{2 m_1^2}J_2(m_1,T)
 \left({\bf C_{12}}-2\ln\left(\frac{m_1^2}{\Lambda^2}\right)\right)+
\\[4mm]\di  +{\bf
C_{13}}\;m_1^2\ln^2\left(\frac{m_1^2}{\Lambda^2}\right) +{\bf
C_{14}}\;m_1^2\ln\left(\frac{m_1^2}{\Lambda^2}\right)+ {\bf
C_{15}}\;m_1^2 \biggr)+(N-1)\biggl(I_2^1(m_2,m_2,m_1,T)+
\\[4mm]\di +K_2^1(m_2,m_2,m_1,T) - \frac12 J_1(m_2,T)+\frac1{2
m_2^2}J_2(m_2,T) \left({\bf
C_{12}}-2\ln\left(\frac{m_2^2}{\Lambda^2}\right)\right)+
\\[4mm]\di +{\bf
C_{16}}\;m_1^2\ln^2\left(\frac{m_2^2}{\Lambda^2}\right) +{\bf
C_{17}} m_2^2\ln^2\left(\frac{m_2^2}{\Lambda^2}\right) +{\bf
C_{18}}\;m_1^2\ln\left(\frac{m_2^2}{\Lambda^2}\right) +{\bf
C_{19}} m_2^2\ln\left(\frac{m_2^2}{\Lambda^2}\right)+ \\[4mm]\di
+{\bf C_{20}}m_1^2+{\bf C_{21}}m_2^2+L^1_2(m_2,m_2,m_1,4/3)
\biggr)\biggr], \end{array} \] \[ \begin{array}{c} \di
\lim_{\1\rightarrow\0}
D^{''}_{_1(\0\1)_{ren}}\equiv\left(\frac1{n}\frac{V_n}{V}\cdot\partial_{i\hspace*{0.5mm}
(x)}\partial^i_{(x_{1})}D_{_1(x
x_1)}\right)_{ren}=-\frac13\cdot\frac{2\la
m_1^2}{N(2\pi)^2}\biggl[ I^1_2(m_1,m_2,m_2,T)+ \\[4mm]\di
+\frac12 K(m_1,m_2,m_2,T)+ \frac12
K(m_2,m_1,m_2,T)-\frac14\frac{m_2}{m_1}J_1(m_1,T)
-\frac14\frac{m_1}{m_2}J_1(m_2,T)+ \\[4mm]\di
\frac14\frac{m_2}{m_1^3}J_2(m_1,T) \left({\bf
C_{12}}-2\ln\left(\frac{m_1^2}{\Lambda^2}\right)\right)
+\frac14\frac{m_1}{m_2^3}J_2(m_2,T) \left({\bf
C_{12}}-2\ln\left(\frac{m_2^2}{\Lambda^2}\right)\right) +
\\[4mm]\di +{\bf
C_{22}}\;m_1^2\ln^2\left(\frac{m_1\;m_2}{\Lambda^2}\right)+ {\bf
C_{23}}\;m_1\;m_2\ln^2\left(\frac{m_1\;m_2}{\Lambda^2}\right)
+{\bf C_{24}}\;m_1^2\ln\left(\frac{m_1\;m_2}{\Lambda^2}\right)+
 \\[4mm]\di
 +{\bf C_{25}}\;m_1\;m_2\ln\left(\frac{m_1\;m_2}{\Lambda^2}\right)
+{\bf C_{26}}\;m_1^2+{\bf C_{27}}\;m_1\;m_2+L^1_2(m_1,m_2,m_2,1)
\biggr] \end{array} \] Numerical values of the coefficients ${\bf
C_{\it i}}$ are: ${\bf C_{12}}\approx 0.719$, ${\bf
C_{13}}=0.011$, ${\bf C_{14}}\approx0.177$, $ {\bf
C_{15}}\approx-0.019$, $ {\bf C_{16}}\approx0.065$, $ {\bf
C_{17}}\approx-0.054$, $ {\bf C_{18}}\approx-0.021$, $ {\bf
C_{19}}\approx0.198$, $ {\bf C_{20}}\approx0.026$, $ {\bf
C_{21}}\approx-0.046$, $ {\bf C_{22}}\approx-0.526$, $ {\bf
C_{23}}\approx0.617$, $ {\bf C_{24}}\approx0.168$, $ {\bf
C_{25}}\approx1.341$, $ {\bf C_{26}}\approx0.396$, $ {\bf
C_{27}}\approx-0.636$.

Thus specific free energy density calculated accurate to first
order of vanishing terms looks like: \[\di\frac{\widetilde{{\cal
F}}}{V}=\lim_{\1\rightarrow\0} G^{''}_{_1(\0\1)_{ren}}+(N-1)\di
\lim_{\1\rightarrow\0} D^{''}_{_1(\0\1)_{ren}}+\frac{m_1^2
v^2}{2}-\frac{4\la v^4}{N}+F(T,m_1(T),m_2(T),v(T)),\] where
(\ref{GenFunc}) is evaluated on solution of (\ref{O(N) m_1}),
(\ref{O(N) m_2}) and (\ref{O(N)condensat_2}).

\paragraph{Numerical calculations and conclusions.}
Taking into account correction to Hartree--Fock approximation does
not bring to appearance of a new branch of solutions of equations
for effective masses and condensate. These values on temperature
dependencies are depicted at Fig. 2 (solid lines at Fig 2.a, Fig
2.b, Fig 2.c). For easier comparison, by dashed line these values
on temperature dependencies in Hartree--Fock approximation are
depicted.\\

\begin{center}
\includegraphics[width=16.5cm]{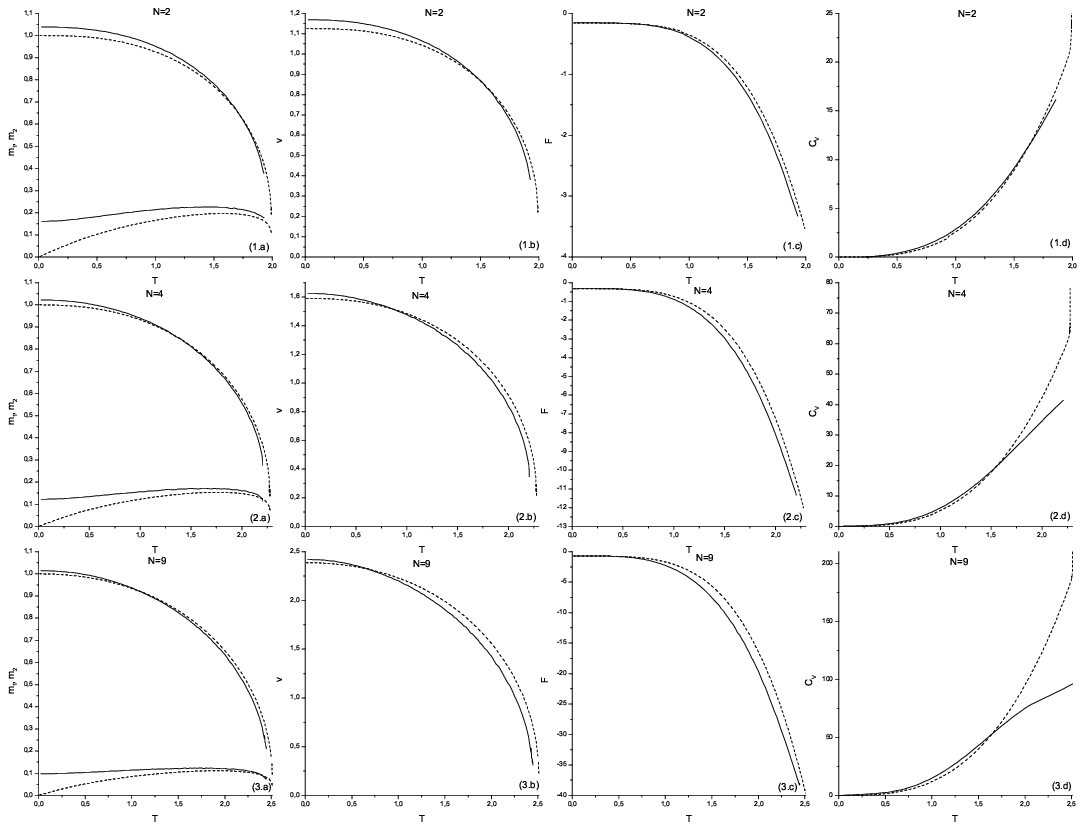}
\\
\rm Fig.2 Effective masses, order parameter, equilibrium free
energy and specific heat capacity on temperature dependencies
calculated accurate to corrections of the first order of
vanishing.
\end{center}

From Fig.2 it follows that in low--symmetry phase only one branch
of solutions remain thermodynamically stable in temperature region
$0\leq T\leq T_{c_2}$.

Significant deviation of $m_2$ on temperature dependence in
compare with that in Hartree--Fock approximation concerns with
mentioned above logarithmically  divergent terms $\di
m_1^2\ln\left(\frac{m_2^2}{\Lambda^2}\right)$ and  $\di
m_1^2\ln\left(\frac{m_1\;m_2}{\Lambda^2}\right)$ in three-point WF
and corrections to two--point WF.

Critical temperature $T_{c_2}$ decreases in compare with that
calculated in Hartree--Fock approximation. With increase of $N$,
relative magnitude of corrections, estimated as quotient $\di
\gamma\equiv\frac{T_{c_2\; PSP}-T_{c_2}}{T_{c_2}}\cdot 100\%$
decreases ($\gamma|_{N=2}\approx3.26\%$,
$\gamma|_{N=4}\approx3.21\%$, $\gamma|_{N=9}\approx2.61\%$) and at
the limit $N\rightarrow\infty$ tends to zero, which agrees with
conclusion made in chapter \ref{Tcepochka}.

Though $m_2$ on temperature dependence significantly changes, at
low temperatures thermodynamical observables (equilibrium free
energy and heat capacity) values are close to those in
Hartree--Fock approximation, as expected from general conclusions.
The closer to $T_{c_2}$, the greater deviation of observables from
that in Hartree--Fock approximation, in particular heat capacity,
which agrees with made above conclusion of formality of heat
capacity divergence at $T_{c_2 (0)}$ vicinity.

\end{document}